\documentclass[letterpaper,twocolumn,10pt]{article}
\usepackage{usenix-2020-09}

\usepackage{titlesec}
\titlespacing*{\paragraph}
{0pt}{1.0ex plus .5ex minus .2ex}{1ex plus .2ex minus .2ex}

\def\BibTeX{{\rm B\kern-.05em{\sc i\kern-.025em b}\kern-.08em
    T\kern-.1667em\lower.7ex\hbox{E}\kern-.125emX}}

\usepackage[nolist]{acronym}

\usepackage{enumitem}

\usepackage{comment}
\usepackage{dirtytalk}

\usepackage{amsmath}
\usepackage{amsfonts}
\usepackage{amssymb}
\usepackage{mathtools}
\usepackage{braket} % multi-row sets

\usepackage{xspace}

\usepackage{balance}

\usepackage{textcomp}

\usepackage[]{subcaption}
\usepackage[font=small]{caption}

\usepackage{booktabs}
\usepackage{multirow}
\usepackage{tabularx}

\newcolumntype{C}[1]{>{\centering\arraybackslash}p{#1}}

\usepackage{multicol}

\usepackage{pifont} % used to get \ding command
\usepackage{wasysym} % more symbols, like circle, rightcircle,...

\usepackage{pgfplots} % External TikZ/PGF support
\usepackage{tikz}
\pgfplotsset{compat=1.13}
\usetikzlibrary{calc}
\usetikzlibrary{positioning,arrows,fit,matrix}
\usetikzlibrary{backgrounds}
\usetikzlibrary{patterns}
\usetikzlibrary{shapes}
\usetikzlibrary{snakes}
\usetikzlibrary{shapes.arrows} % used for wide arrow only
\usetikzlibrary{chains}
\usepgfplotslibrary{groupplots}
\usepgfplotslibrary{statistics}
\usepgfplotslibrary{fillbetween}

\definecolor{commentgreen}{HTML}{316D94} % 
\definecolor{cbone}  {HTML}{006BA4} % 
\colorlet{primarycolor}{cbone}

\usepackage{cleveref} % 
\usepackage[numbers,sort]{natbib} 

\usepackage{microtype}

\newcommand{\ie}{i.\,e.}
\newcommand{\eg}{e.\,g.}

\newcommand{\tabspac}{\addlinespace[4pt]}

\newcommand{\appendixref}[1]{\hyperref[#1]{Appendix~\ref{#1}}}

\newcommand{\cmark}{\ding{51}}%
\newcommand{\xmark}{}%

\newcommand{\stepone}{\ding{182}\xspace}
\newcommand{\steptwo}{\ding{183}\xspace}
\newcommand{\stepthree}{\ding{184}\xspace}
\newcommand{\stepfour}{\ding{185}\xspace}
\newcommand{\stepfive}{\ding{186}\xspace}
\newcommand{\stepsix}{\ding{187}\xspace}
\newcommand{\stepseven}{\ding{188}\xspace}

\hypersetup{
    colorlinks = false,
    hidelinks = true
}

\newcommand{\bow}{\textbf{x}}
\newcommand{\vocabulary}{\mathcal{V}}

\newcommand{\assignment}{\mathsf{A}}

\newcommand{\corpus}{D}
\newcommand{\corpusdoc}{\bow}
\newcommand{\corpusdoclen}{\left|\corpusdoc\right|}

\newcommand{\topicwordprior}{\beta}
\newcommand{\topicdocumentprior}{\alpha}

\newcommand{\submission}{\bow}
\newcommand{\submissions}{\mathcal{S}}

\newcommand{\reviewersset}{\mathcal{R}}
\newcommand{\reviewersubset}{R}
\newcommand{\surroundingreviewers}{R}
\newcommand{\reviewer}{r}
\newcommand{\archive}{A}

\newcommand{\reviewerload}{\mathsf{L}_{\reviewer}}
\newcommand{\paperload}{\mathsf{L}_{\submission}}

\newcommand{\select}{\text{sel}}
\newcommand{\reject}{\text{rej}}

\newcommand{\requestedreviewers}{\reviewersubset_{\select}}
\newcommand{\rejectedreviewers}{\reviewersubset_{\reject}}

\newcommand{\bid}{b}

\newcommand{\reviewerwindow}{\omega}
\newcommand{\revieweroffset}{\upsilon}
\newcommand{\reviewerwordsmax}{\nu}

\newcommand{\reviewerwordsmass}{Q}
\newcommand{\reviewerwords}{\hat{\topicworddist}}

\newcommand{\topic}{t}
\newcommand{\topics}{T}

\newcommand{\topicspace}{\mathcal{T}}

\newcommand{\stepsize}{k}
\newcommand{\nosuccessors}{M}

\newcommand{\loss}{\ell}

\newcommand{\extractor}{\Phi}
\newcommand{\topicextractor}{\Gamma}
\newcommand{\topicworddist}{\phi}

\newcommand{\topicdocumentdist}{\theta}

\newcommand{\word}{w}

\newcommand{\prob}{P}

\newcommand{\beamwidth}{B}
\newcommand{\maxitr}{I}

\newcommand{\switches}{S}

\newcommand{\attackbudgetscale}{\sigma}

\newcommand{\modifications}{\delta}
\newcommand{\modificationsmannorm}{{\left| \left| \modifications \right| \right|}_1}
\newcommand{\modificationsinfnorm}{{\left| \left| \modifications \right| \right|}_\infty}

\newcommand{\maxmannorm}{L_1^\text{max}}
\newcommand{\maxinfnorm}{L_\infty^\text{max}}

\newcommand{\margin}{\gamma}

\newcommand{\Dom}{\ensuremath{\mathcal{Z}}\xspace}
\newcommand{\F}{\ensuremath{\mathcal{F}}\xspace}
\newcommand{\preprocessing}{\ensuremath{\rho}\xspace}
\newcommand{\inputpdf}{\ensuremath{z}\xspace}

\newcommand{\transformation}{\ensuremath{\omega}\xspace}
\newcommand{\transformations}{\ensuremath{\Omega}\xspace}
\begin{acronym}
    \acro{LDA}{\emph{Latent Dirichlet Allocation}}
    \acro{CMT}{\emph{Conference Management Toolkit}}
    \acro{TPMS}{\emph{The Toronto Paper Matching System}}
    \acro{MCMF}{\emph{MinCost-MaxFlow}}
\end{acronym}

\begin{document}
\date{}

\title{\Large \bf No more Reviewer \#2: Subverting Automatic  Paper-Reviewer \\ Assignment using Adversarial Learning}

\author{
{\rm 
    Thorsten Eisenhofer\thanks{Shared first authorship}~\footnotemark[2]~,~%
    Erwin~Quiring\footnotemark[1]~\footnotemark[2]~\footnotemark[3]~,~%
    Jonas Möller\footnotemark[4]~,~%
    Doreen Riepel\footnotemark[2]~,~%
  }
  \\
  {\rm
    Thorsten Holz\footnotemark[5]~,~%
    Konrad~Rieck\footnotemark[4]\vspace{0.3em}
  }\\
{\normalsize \footnotemark[2]~~Ruhr University Bochum}\\
{\normalsize \footnotemark[3]~~International Computer Science Institute (ICSI) Berkeley}\\
{\normalsize \footnotemark[4]~~Technische Universität Berlin}\\
{\normalsize \footnotemark[5]~~CISPA Helmholtz Center for Information Security}\vspace{1em}\\ 
}

\maketitle

\begin{abstract}
The number of papers submitted to academic conferences is steadily rising in many scientific disciplines. To handle this growth, systems for automatic \emph{paper-reviewer assignments} are increasingly used during the reviewing process. These systems use statistical topic models to characterize the content of submissions and automate the assignment to reviewers. In this paper, we show that this automation can be manipulated using adversarial learning. We propose an attack that adapts a given paper so that it misleads the assignment and selects its own reviewers. Our attack is based on a novel optimization strategy that alternates between the feature space and problem space to realize unobtrusive changes to the paper. To evaluate the feasibility of our attack, we simulate the paper-reviewer assignment of an actual security conference (IEEE S\&P) with 165 reviewers on the program committee. Our results show that we can successfully select and remove reviewers without access to the assignment system. Moreover, we demonstrate that the manipulated papers remain plausible and are often indistinguishable from benign submissions. 
\end{abstract}
\vspace{-0.5em}
\section{Introduction}
\label{sec:introduction}

Peer review is a major pillar of academic research and the scientific publication process. Despite its well-known weaknesses, it is still an essential instrument for ensuring high-quality standards through the independent evaluation of scientific findings~\cite{soneji-22-experts, misc-nips, misc-esa}. For this evaluation, a \emph{submission} is assigned to a group of reviewers, taking into account their expertise, preferences, and potential biases. For conferences, this assignment is traditionally carried out by a program chair, while for journals, the task is performed by an editor. This mechanism has proven effective in the past, but is becoming increasingly difficult to realize as research communities grow. For example, the number of papers submitted to top-tier security conferences is increasing exponentially, reaching over 3,000 submissions in 2020. Likewise, the number of reviewers continuously grows for all major security conferences~\citep{misc-circus}.

To handle this growth, conference management tools have become indispensable in peer review. They allow reviewers to bid for submissions and support the program chair to find a good assignment based on a best-effort matching. Unfortunately, even these tools reach their limit when the number of submissions continues to grow and manual bidding becomes intractable, as for example, in the area of machine learning. Major conferences in this area regularly have over 10,000 submissions that need to be distributed among more than 7,000 reviewers~\cite{misc-neurips}. For this reason, conference management tools are increasingly extended with automatic systems for \emph{paper-reviewer assignment}~\cite{misc-autobid, charlin-13-toronto}. These systems use topic models from machine learning to assess reviewer expertise, filter submissions, and automate the assignment process. 

In this work, we show that this automation can be exploited to manipulate the assignment of reviewers. In contrast to prior work that focused on bid manipulations and reviewer collusion~\citep{jecmen-20-mitigating, wu-21-making}, our attack rests on adversarial learning. In particular, we propose an attack that adapts a given paper so that it misleads the underlying topic model. This enables us to reject and select specific reviewers from the program committee. To reach this goal, we introduce a novel optimization strategy that alternates between the feature space and problem space when adapting a paper. This optimization allows us to preserve the semantics and plausibility of the document, while carefully changing the assignment of reviewers.

Our attack consists of two alternating steps: 
First, we aim at misleading the topic model employed in current assignment systems~\cite{charlin-13-toronto, misc-autobid}. This model defines a latent topic space that is difficult to attack because neither gradients nor an explicit decision boundary exist. To address this problem, we develop a search algorithm for exploring the latent space and manipulating decisions in it.
As a counterpart, we introduce a framework for modifying papers in the problem space. This framework provides several transformations for adapting the paper's content, ranging from invisible comments to synonym replacement and generated text. These transformations enable us to preserve the paper's semantics, while gradually changing the assignment of reviewers.

To empirically evaluate the practical feasibility of our attack, we simulate the paper-reviewer assignment of the 43rd IEEE Symposium on Security and Privacy (IEEE S\&P) with the original program committee of 165 reviewers in both a \emph{black-box} and a \emph{white-box} threat scenario. As the basis for our attacks, we consider 32 original submissions that are publicly available with \LaTeX{} source code. 

Our white-box adversary achieves an alarming performance: we can successfully remove \emph{any} of the initially assigned reviewers from a submission, and even scale the attack to completely choose \emph{all} reviewers in the automated assignment process. 
In the black-box scenario, we can craft adversarial papers that transfer to an unknown target system by only using public knowledge about a conference. We achieve a success rate of up to 90\% to select a reviewer and 81\% to reject one. Furthermore, we demonstrate that the attack remains robust against variations in the training data.

Our work points to a serious problem in the current peer review process: With the application of machine learning, the process inherits vulnerabilities and becomes susceptible to new forms of manipulation.
We discuss potential defenses: (1) For the feature space, robust topic modeling may limit the attacker's capabilities and (2) for the problem space, we recommend using optical character recognition (OCR) techniques to retrieve the displayed text.
Nevertheless, these safeguards cannot completely fend off our manipulations and reviewers should be made aware of this threat.

\paragraph{Contributions.}
We make the following key contributions:
\begin{itemize}[topsep=3pt, itemsep=3pt, partopsep=3pt, parsep=3pt]
\item \emph{Attack against topic models.}  We introduce a novel attack against topic models suitable for manipulating the ranking of reviewers. The attack does not depend on the availability of gradients and explores the latent topic space through an efficient beam search.
\item \emph{Problem-space transformations.} Our attack ensures that both the semantics and plausibility of the generated adversarial papers are preserved. This goal is achieved by a variety of transformations that carefully manipulate the document format and text of a submission.
\item \emph{Adversarial papers.} 
We present a method for constructing adversarial papers in a black-box and white-box scenario, unveiling a serious problem in automatic reviewer assignment. The attack rests on a novel hybrid approach to construct adversarial examples in discrete domains
\end{itemize}
Examples of the created adversarial papers are provided at \href{https://github.com/rub-syssec/adversarial-papers}{https://github.com/rub-syssec/adversarial-papers}. We also make our code and artifacts publicly available here.

\section{Technical Background}
\label{sec:background}

Let us start by reviewing the necessary background for the design of our attack, covering the process of paper-review assignment and the underlying topic modeling. 

\paragraph{Systems for paper-reviewer assignment.} 
To cope with the abundance of submissions, several concepts have been proposed to assign reviewers to submissions~\citep[e.g.,][]{liu-14-robust, li-13-automatic, stelmakh-19-peerreview, long-13-good}. In practice, the most widely used concept is \ac{TPMS} by \citet{charlin-13-toronto}. Because of its high-quality assignments and direct integration with Microsoft's conference management tool CMT~\cite{misc-cmt}, \ac{TPMS} is used by numerous conferences in different fields, including ACM CCS in 2017--2019 and NeurIPS/ICML. \ac{TPMS} can be considered the de facto standard for automatic matching of papers to reviewers. Unfortunately, the implementation of \ac{TPMS} is not publicly available and thus we focus in this work on \emph{Autobid}~\cite{misc-autobid}, an open-source realization of the \ac{TPMS} concept. 
Autobid closely follows the process described by \citet{charlin-13-toronto}. The system has been designed to work alongside HotCRP~\cite{misc-hotcrp} and was used to support reviewer assignment at the IEEE Symposium on Security and Privacy (S\&P) in 2017 and 2018.

Technically, \ac{TPMS} and Autobid implement a processing pipeline similar to most matching concepts: (a) the text from the submission document is extracted and cleansed using natural language processing, (b) the preprocessed text is then mapped to the latent space of a topic model, and finally (c) an assignment is determined by deriving a ranking of reviewers. In the following, we review these steps in detail.

\begin{figure}[b]
    \centering
        \includegraphics{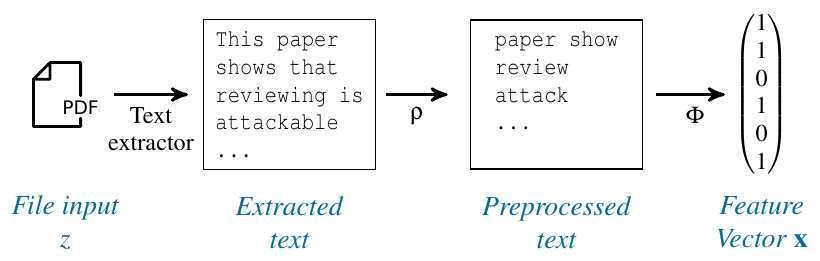}
    \caption{Text preprocessing in paper-reviewer assignment.}
    \label{fig:background_pipeline}
\end{figure}

\paragraph{(a) Text preprocessing.}
When working with natural languages, multiple steps are required to bring text into a form suitable for machine learning (see Figure~\ref{fig:background_pipeline}). 
As paper submissions can be provided in different formats, the pipeline starts by extracting text from the underlying document, typically the PDF format. This original document resides in the problem space of our attack and is denoted as \inputpdf in the following. Autobid employs the tool \texttt{pdftotext} for this task, which is used in our evaluation in Section~\ref{sec:evaluation}.
The extracted text is then normalized using a preprocessor function \preprocessing. Typically, it is tokenized, converted to lowercase, and stemmed~\cite{lovins-68-development}. Subsequently, stop words are removed so that each submission is now represented as a sequence of filtered stems. Autobid employs the NLTK package \citep{bird-09-natural} to perform this task. 

Finally, a feature extractor $\extractor$ maps the input $\preprocessing(\inputpdf)$ to a bag-of-words vector $\bow \in \mathbb{N}^{|\vocabulary|}$ with $\vocabulary$ being the vocabulary formed over all words (stems). That is, a submission is represented by a high-dimensional vector whose individual dimensions reflect the count of words.  Although this representation is frequently applied in supervised learning, the high dimensionality is problematic for unsupervised learning and complicates determining topics in the submission.

\paragraph{(b) Topic modeling.}
The key to matching reviewers to papers is the automatic identification of topics in the text. This unsupervised learning task is denoted as \emph{topic modeling}. While there exist several algorithms for this modeling, many assignment systems, including \ac{TPMS} and Autobid, use \ac{LDA}. 
\ac{LDA} is a Bayesian probabilistic method for topic modeling that allows representing a document as a low-dimension mixture of latent topics. 
Formally, we define this representation as a function
\[ 
   \topicextractor\colon \mathbb{N}^{\left|\vocabulary\right|} \longrightarrow \topicspace, \quad 
   \bow\mapsto \topicdocumentdist_\bow
\]
mapping a bag-of-words vector $\bow$ to a low-dimensional vector space $\topicspace$, whose dimensions reflect different  topics.

Generally, \ac{LDA} is modeled as a generative probabilistic process \cite{blei-02-lda, hoffman-10-online, darling-11-theoretical}. It assumes a corpus $\corpus$ of documents and models each document as a random mixture over a set of latent topics $\topics$. A topic $\topic$ is characterized by a multinomial distribution over the vocabulary $\vocabulary$, and drawn from a Dirichlet distribution $\topicworddist_\topic \sim \textit{Dirichlet}(\topicwordprior)$ with the prior $\topicwordprior$. The Dirichlet prior is usually sparse (i.e., $\topicwordprior<1$) to model that words are not uniformly assigned to topics. Given these topics, for each document $\corpusdoc \in \corpus$, a distribution of topics $\topicdocumentdist_\corpusdoc \sim \textit{Dirichlet}(\topicdocumentprior)$ is drawn. Again, the prior $\topicdocumentprior$ is sparse to account for that documents are usually only associated with a small number of topics. Finally, for each word $\word_i \in \corpusdoc$, a topic $\topic_i \sim \textit{Multinom}(\topicdocumentdist_\corpusdoc)$ is selected and the observed word $\word_i \sim \textit{Multinom}(\topicworddist_{\topic_i})$ is drawn. 
This process can be summarized by the joint probability
\begin{equation}
    \prob(\mathbf{\word},\mathbf{\topic},\topicdocumentdist,\topicworddist|\topicdocumentprior, \topicwordprior) = \prob(\topicworddist|\topicwordprior)\prob(\topicdocumentdist|\topicdocumentprior)\prob(\mathbf{\topic}|\topicdocumentdist)\prob(\mathbf{\word}|\topicworddist_{\mathbf{\topic}})
\end{equation}
with $\mathbf{\word} = (\word_1, \dots, \word_{\corpusdoclen}) $ and $\mathbf{\topic} = (\topic_1, \dots, \topic_{\corpusdoclen})$.

To create a topic model in practice, we need to reverse this process and learn the posterior distribution of the latent variables $\mathbf{\topic}$, $\topicdocumentdist$, and $\topicworddist$ given the \emph{observed} documents $\corpus$. Specifically, we need to solve
\begin{equation}
    \prob(\topicdocumentdist, \topicworddist, \mathbf{\topic}|\mathbf{\word}, \topicdocumentprior, \topicwordprior) = \frac{\prob(\topicdocumentdist, \topicworddist, \mathbf{\topic}, \mathbf{\word}|\topicdocumentprior, \topicwordprior)}{\prob(\mathbf{\word}|\topicdocumentprior, \topicwordprior)}.
\end{equation}
Solving this equation is intractable as the  term $\prob(\mathbf{\word}|\topicdocumentprior, \topicwordprior)$ cannot be computed exactly~\cite{blei-02-lda}. To address this, different approximated techniques, such as variational inference \cite{blei-02-lda, hoffman-10-online} or Gibbs Sampling \cite{darling-11-theoretical}, are typically used for implementations of \ac{LDA}. 
Autobid builds on variational inference based on the implementation of GenSim~\cite{rehurek-10-gensim}.

For the feature vector $\corpusdoc$ of a new submission, the same technique---conditioned on the corpus $\corpus$---is used to compute the corresponding topic mixture $\topicdocumentdist_\corpusdoc$. Attacking this process is challenging, as no gradients or other guides for moving in the direction of particular topics are directly available. Hence, we develop a new search algorithm for subverting the topic assignment of \ac{LDA} in Section~\ref{sec:feature-space}.

\paragraph{(c) Paper-reviewer assignment.}
\label{sec:background:assignment}
Finally, the topic model is used to estimate the reviewer expertise and automate the matching of submissions to reviewers. More specifically, let $\reviewersset$ be the set of all potential reviewers and $\submissions$ a set of submissions $\submission \in \mathbb{N}^{|\vocabulary|}$. For each reviewer $\reviewer$, we collect an archive $\archive_\reviewer \in \mathbb{N}^{|\vocabulary|}$ representative of the reviewer's expertise and interests. Since researchers are naturally described best by their works, this could, for example, be a selected set of previously published papers. The corresponding archives are modeled as a union over all papers.

For each pair of reviewer $\reviewer$ and submission $\submission$, a \emph{bidding score} $\bid_{\reviewer, \submission}$ is calculated. This score reflects the similarity between the reviewer's archive $\archive_\reviewer$ and a submission $\submission$: the more similar, the higher the score. Given the topic extractor $\topicextractor(\cdot)$, a reviewer $\reviewer$ and a submission $\submission$, Autobid defines the bidding score as the following dot-product
\begin{equation}
\bid_{\reviewer,\submission} \coloneqq \topicextractor(\archive_\reviewer) \cdot \topicextractor(\submission)^\top.
\end{equation}

Subsequently, these bidding scores are used for the final assignment $\assignment \in \{0,1\}^{|\reviewersset|\times|\submissions|}$ with the goal to maximize the similarity between reviewers and submissions.
In this phase, additional constraints are included: the assignment is subjected to (1) the targeted number of reviewers $\paperload$ assigned to a submission and (2) the maximum number of submissions $\reviewerload$ assigned to a reviewer.
More formally, we can describe the assignment as the following bipartite matching problem:

\begin{equation*}
\begin{array}{ll@{}ll}
\underset{\scriptscriptstyle \assignment}{\text{maximize}}    & \displaystyle \displaystyle\sum_{\reviewer}\sum_{\submission} \bid_{\reviewer, \submission} \cdot \assignment_{\reviewer, \submission} &\\[3ex]
\text{subject to}   & \assignment_{\reviewer, \submission} \in \{0,1\} \;\forall \reviewer,\submission \\[2ex]
                    & \displaystyle \sum_{\reviewer} \assignment_{\reviewer, \submission} \leq \paperload  \;\forall \submission \\[3ex]
                    & \displaystyle \sum_{\submission} \assignment_{\reviewer, \submission} \leq \reviewerload \;\forall \reviewer \\
                    
\end{array}
\end{equation*}
This optimization problem can then be reformulated and efficiently solved with \emph{Linear Programming (LP)} \cite{taylor-08-optimal}.
\section{Adversarial Papers}
\label{sec:approach}
\begin{figure*}[t]
    \centering
    % l b r t
  	\includegraphics[trim=0 0 0 0, clip, width=0.8\textwidth]{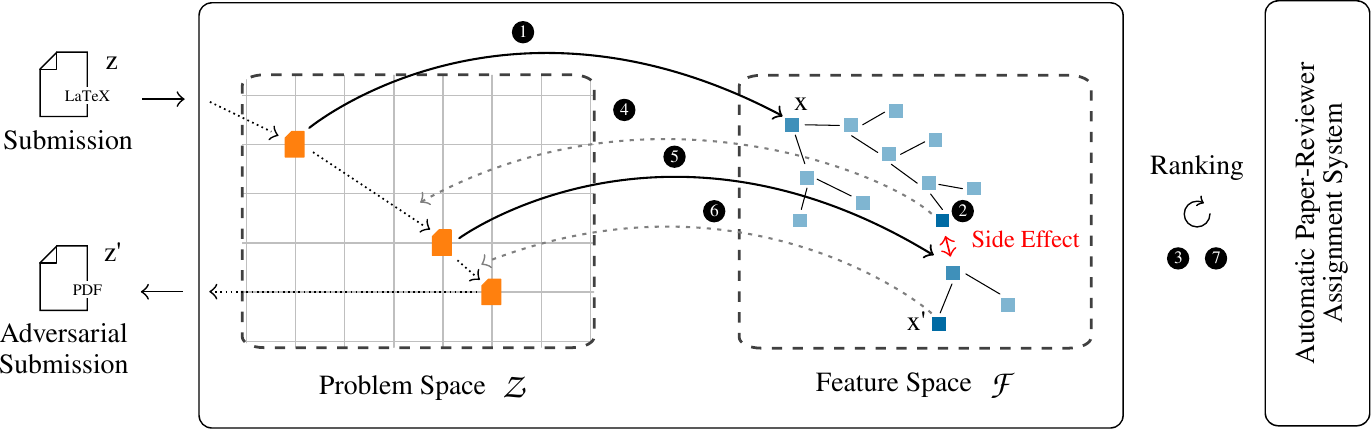}
    \caption{\textbf{Feature-problem-space attack.} For a submission $\inputpdf$, we construct an adversarial submission~$\inputpdf'$ that leads to a targeted assignment of reviewers. Our attack alternately switches between \Dom and \F. In step~\stepone, we extract word counts $\submission$ from submission $\inputpdf$, and use a search algorithm to change $\submission$ in~\F to obtain the desired ranking (step~\steptwo). To guide this search, we query the paper-reviewer assignment system for scores (step~\stepthree). Next, we realize the modifications in the problem space~\Dom and manipulate \inputpdf (step~\stepfour). Projecting the resulting submission back to \F, the submission vector will be shifted due to side effects and transformation limitations (step~\stepfive). This shift is considered by continuing the search process from this new position and repeating this process iteratively (step~\stepsix) until we obtain a valid adversarial submission $\inputpdf'$ (step~\stepseven).}
    \label{fig:overview}
  	% \vspace{-0.75em}
\end{figure*}

We proceed to introduce our approach for subverting the paper-reviewer assignments. To this end, we first define a threat model for our attack, and then examine challenges and required steps to control the matching.

\paragraph{Threat model.}
We consider a scenario where the adversary only modifies her submission---the \emph{adversarial paper}---to manipulate the assigned reviewers. We assume two representative classes of adversaries with varying degrees of knowledge.
First, we focus on \emph{white-box adversaries} with complete access to the assignment system, including the trained model and reviewer archives. This represents a very strong class of adversaries and allows us to generally study the strength as well as limitations of our attack against assignment systems.
Second, we study the more realistic scenario with a \emph{black-box adversary}. The adversary is assumed to have only a general knowledge about the assignment system (\ie, AutoBid is an open-source project~\cite{misc-autobid}). No access to the training data and learned model is given. In this setting, we envision adversaries that exploit public information about a conference, such as knowledge about the program committee.

\paragraph{Challenges.}
The adversary has to operate both in the problem space~\Dom and the feature space~\F. The former consists of the input objects (\eg{}, the \LaTeX{} source files of the paper); the latter contains the feature vectors that are used as inputs for the learning system. In contrast to domains like image recognition, the mapping from the problem space to the feature space is not bijective, \ie{}, there is no one-to-one correspondence between \Dom and \F. This poses a challenge for the adversary because a promising feature vector may not be mapped to a valid submission. A further obstacle is that some modifications in the problem space cannot be applied without side effects: If an adversary, for instance, adds a sentence  to include a particular word, she inevitably adds other words that change the feature vector. 

To deal with these challenges, we introduce a \emph{hybrid} optimization strategy that alternates between the feature-space and problem-space representations of the attack submission.
This optimization enables us to preserve the semantics and plausibility of the document, while at the same time gradually changing the assignment of reviewers. A general overview of our attack is outlined in Figure \ref{fig:overview}.
Second, we transfer problem-space restrictions to the feature space. In this way, we resolve restrictions in a generic manner without adjusting our problem-space transformations.

\paragraph{Attack goals.}
Given a submission $\inputpdf$, our goal is to find an adversarial paper~$\inputpdf'$ that leads to the adversary's targeted review assignment. In the feature space, we thus want to manipulate the set of assigned reviewers $\reviewersubset_\submission$. That is, we want to \emph{select} and \emph{reject} arbitrary reviewers to be included in respectively excluded from $\reviewersubset_\submission$. Formally, we define two sets $\requestedreviewers$ and $\rejectedreviewers$ and our goal is to find a vector $\modifications \in \F$ such that the modified word counts $\submission' \coloneqq \submission + \modifications$ fulfill
\begin{equation}
\label{eq:feature-space-attack-goal}
\begin{split}
    \reviewer \in \requestedreviewers & \Rightarrow r \in \reviewersubset_{\submission'} \text{, and} \\
    r \in \rejectedreviewers & \Rightarrow r \notin \reviewersubset_{\submission'} , \forall \reviewer \in \reviewersset .
\end{split}
\end{equation}
We require every reviewer $\reviewer \in \requestedreviewers$ is included in $\reviewersubset_{\submission'}$ and likewise every reviewer $\reviewer \in \rejectedreviewers$ excluded from~$\reviewersubset_{\submission'}$. 
In addition, we take care that the targeted solution is feasible with $ \left| \requestedreviewers \right| \leq \paperload$ and $ \left| \rejectedreviewers \right|  \leq \left| \reviewersset  \right| - \paperload$.

Furthermore, we restrict the modifications to $\modificationsmannorm \leq \maxmannorm$ and $\modificationsinfnorm \leq \maxinfnorm$. The $L_1$ constraint limits the amount of modifications to the submissions and makes the attack less suspicious. Similarly, the $L_\infty$ constraint restricts the maximum change to a single feature, so that a word is not included too frequently. 
Finally, with respect to the concrete assignment process, we assume an automatic matching that always selects the reviewers with the highest assignment scores. We note that this assumption can be relaxed, as shown by \citet{jecmen-20-mitigating}, and combined with colluding reviewers. We further discuss the impact of concurring submissions in Section \ref{sec:discussion}.

For manipulations in the problem space, we design various transformations for adapting the submission~\inputpdf. We denote a single transformation by $\transformation : \Dom \longrightarrow \Dom$,$\;$ $\inputpdf \mapsto \inputpdf'$, where multiple transformations can be chained together as $\transformations = \transformation_k \circ \dots \circ \transformation_2 \circ \transformation_1$. To avoid transformations from creating artifacts and visible clues, we introduce the following problem-space constraints:
First, we need to \emph{preserve the semantics} of the text, so that the paper is still a meaningful submission. Second, we add a \emph{plausibility} constraint, that is, the modifications should be as inconspicuous as possible. 
We summarize the constraints as $\Upsilon$ and write \mbox{$\transformations(\inputpdf) \models \Upsilon$} if a transformation sequence \transformations on a submission fulfills these constraints.

\paragraph{Optimization problem.}
We arrive at the following optimization problem for generating adversarial examples, integrating constraints from the problem space and the feature space:
\begin{equation}
\label{eq:attack-goal}
\begin{split}
  & r \in \requestedreviewers \Rightarrow r \in \reviewersubset_{\submission'} \text{, and} \\ 
  &  r \in \rejectedreviewers \Rightarrow r \notin \reviewersubset_{\submission'}, \forall \reviewer \in \reviewersset\\
\text{ subject to \;\; } & 
\modificationsmannorm \leq \maxmannorm \text{ and }
 \modificationsinfnorm \leq \maxinfnorm \\
& \transformations(\inputpdf) \models \Upsilon 
\end{split}
\end{equation}
with \mbox{$\submission \; = \extractor \circ \preprocessing (\inputpdf)$}, \mbox{$\submission' = \extractor \circ \preprocessing (\transformations(\inputpdf))$}, and \mbox{$\modifications = (\submission' - \submission)$}.
We proceed to realize this optimization strategy by first introducing our attack in the feature space and then in the problem space, before merging both components.

\subsection{Feature Space}
\label{sec:feature-space}

In an automatic paper-reviewer assignment system, the set of reviewers $\reviewersubset_\submission$ for a submission is determined by the computed assignment scores $\bid_{\reviewer, \submission}$ between reviewers $\reviewer$ (characterized by their archives $\archive_\reviewer$) and the submission vector $\submission$: 
\begin{equation}
    \label{eq:score}
    \bid_{\reviewer, \submission}  \coloneqq \topicextractor(\archive_\reviewer) \cdot \topicextractor(\submission)^\top
\end{equation}

To change the assignment score and thus affect the matching, we can only influence the extracted high-level features $\topicextractor(\submission)$ since $\archive_r$ is fixed for a given set of $\reviewersset$. However, even when we have full control over $\topicextractor(\submission)$, changing the relative ordering---the \emph{ranking}---between reviewers is not straightforward. For instance, suppose we have two reviewers $\reviewer_1$ and $\reviewer_2$ that share most topics (i.e., $\topicextractor(\archive_{\reviewer_1}) \approx \topicextractor(\archive_{\reviewer_2})$), adjusting $\topicextractor(\submission)$ in this case will have a similar effect on both.
In particular, if we na\"ively  try to increase the assignment score from $\reviewer_1$, we simultaneously also increase the score of $\reviewer_2$ and vice versa. Even if reviewers are not working in the same area, their topic distributions often partially overlap, as their research builds on similar principles and concepts. Hence, to modify the ranking we need to carefully maneuver the submission in the feature space.
This is significantly more challenging compared to attacking a classification, as we need to both attack the model's prediction while simultaneously considering effects on concurring reviewers.

Our attack is further complicated by the fact that altering the topic distribution itself is a challenging task, since we need to make changes in the latent topic space. For \ac{LDA}, this distribution $\topicextractor(\submission) = \topicdocumentdist_\submission$ is computed using a probabilistic inference procedure. Thus, typical gradient-style attacks are not applicable. Indeed, \citet{zhou-20-evalda} even show that the manipulation of this inference is \emph{NP-hard}. Moreover, \ac{LDA} typically assigns only a small weight to individual words, so an attacker is required to manipulate a comparatively large set of words for subverting the topic assignment.

To address both of these challenges, we use a stochastic beam search. For a given submission vector $\submission$, we start with an empty modification vector $\modifications$ which is extended in each iteration until we find a successful submission vector $\submission' \coloneqq \submission + \modifications$ or a maximum number of iteration $\maxitr$ is reached. During this search, we consider $B$ candidate vectors in parallel and select successors after each iteration with a probability increasing as a function of the candidates' loss. 

\paragraph{Loss.}
For our search, we define the following loss function to evaluate the quality of a submission $\submission$ in terms of the objective from Equation \ref{eq:feature-space-attack-goal} that incorporates the selection and rejection of reviewers:
\begin{equation}
    \loss \coloneqq \loss_{\select} + \loss_{\reject}
\end{equation}
For selected reviewers, the loss $\loss_{\select}$ is reduced when the assignment scores $\bid_{\hat{\reviewer}, \submission}$ increase or when the ranks of the reviewers improve (\ie{}, when reviewers ascend in the ranking):
\begin{equation}
\loss_{\select} \coloneqq \sum_{\hat{\reviewer} \in \requestedreviewers}{\text{rank}_\submission^{\hat{\reviewer}} \cdot (1 - \bid_{\hat{\reviewer}, \submission})}
\end{equation}
where $\text{rank}_\submission^{\hat{\reviewer}}$ is the rank of reviewer $\hat{\reviewer}$ for submission $\submission$.
Similarly, for rejected reviewers the loss $\loss_{\reject}$ is reduced when the assignment scores $\bid_{\check{\reviewer}, \submission}$ decrease:
\begin{equation}
\loss_{\reject} \coloneqq \sum_{\check{\reviewer} \in \rejectedreviewers}{\max(\text{rank}^{\reject}_{\submission} - \text{rank}_\submission^{\check{\reviewer}}, 0) \cdot (\bid_{\check{\reviewer}, \submission} - \bid_{\reviewer_{\reject}, \submission})}
\end{equation}
where $\text{rank}^{\reject}_{\submission}$ is the target rank for a rejected reviewer (\ie{}, rejected reviewer are pushed down towards this rank) and $\bid_{\reviewer_{\reject}, \submission}$ is the corresponding assignment score.
This loss is designed to focus on reviewers that are far off, but simultaneously allows reviewers to provide \say{room} for following reviewers, for example, when we want to move a group of reviewers upwards/downwards in the ranking.

We consider a submission vector $\submission$ successful when the objective from Equation \ref{eq:feature-space-attack-goal} is fulfilled. At this point, we are naturally just at the boundary of the underlying decision function. To make the submission vector more \emph{resilient}, we could continue to decrease the loss. However, since we already successfully ordered the reviewer (\ie{}, the ranking), we are more interested in maximizing the \emph{margin} of selected and rejected reviewers to the border of $\reviewersubset_\submission$. We denote this margin as $\margin$ and set $\loss \coloneqq -\margin$ whenever $\submission$ satisfies Equation \ref{eq:feature-space-attack-goal}. Decreasing the loss is then equivalent to maximizing $\margin$. 

\paragraph{Candidate generation.}

\begin{figure}[t]
    \centering
    % l b r t
  	\includegraphics[trim=0 0 0 0, clip, width=0.8\columnwidth]{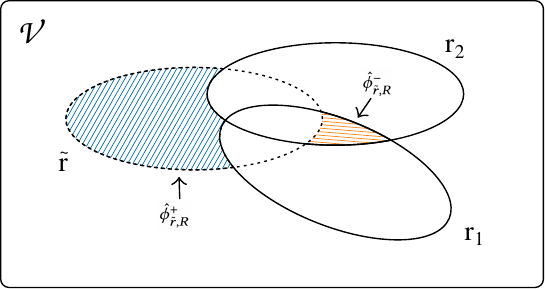}
    \caption{\textbf{Reviewer words.} Based on the topic model, we associate reviewer with a set of predictive words. Given target reviewer $\tilde{\reviewer}$ and a set of concurring reviewers $\surroundingreviewers = \{\reviewer_1, \reviewer_2\}$, we construct distributions $\reviewerwords^+_{\tilde{\reviewer}, \surroundingreviewers}$ and $\reviewerwords^-_{\tilde{\reviewer}, \surroundingreviewers}$. Sampling from these distributions yield words that are only predictive for reviewer $\tilde{\reviewer}$ respectively reviewers in $\surroundingreviewers$. }
    \label{fig:reviewer-words}
  	\vspace{-0.5em}
\end{figure}

A key operation of the beam search is the generation of new candidate vectors. We create a successor from a given submission by adding (respectively removing) $\stepsize$ words to adjust topic distribution $\topicdocumentdist_\submission = \topicextractor(\submission)$ and ultimately the ranking of submission $\submission$.
To select words, we represent (broadly speaking) each reviewer by a set of \emph{predictive} words and sample words that lie in the disjunction between a target and its concurring reviewers. An example of this is shown in Figure \ref{fig:reviewer-words}.

To construct these sets, we first represent each reviewer $\reviewer$ by a  \emph{reviewer to words} distribution $\reviewerwords_\reviewer$ over vocabulary $\vocabulary$. 
Intuitively, this distribution assigns each word the probability how predictive it is for $\reviewer$. Formally, we define the probability mass function for $\reviewerwords_\reviewer$ as follows:
\begin{equation*}
\label{eq:reviewerwords}
   \reviewerwordsmass_\reviewer \colon \vocabulary \rightarrow \mathbb{R}, \quad 
   \word \mapsto 
   \frac
   {\frac{1}{\mid\topics\mid} \sum_{\topic \in \topics}{\prob(\word \mid \topic)\ \prob(\topic \mid \reviewer)}}
   {\sum_{\word\in\vocabulary}{\frac{1}{\mid\topics\mid} \sum_{\topic \in \topics}{\prob(\word \mid \topic)\ \prob(\topic \mid \reviewer)}}}
\end{equation*}

Remember that each topic $\topic$ defines a distribution over $\vocabulary$ and each reviewer can be represented by $\topicextractor(\archive_\reviewer)$. $\reviewerwordsmass_\reviewer$ assigns each word the average probability over all topics $\topics$ scaled by the relevance of topic $\topic$ for reviewer $\reviewer$. 
Randomly sampling from $\reviewerwords_\reviewer$ thus yield words with a probability given as a function of their \emph{predictiveness} for $\reviewer$. In practice, $\vocabulary$ is typically large and most words are assigned with an insignificant probability. To improve performance, we therefore restrict $\reviewerwords_\reviewer$ to the $\reviewerwordsmax$ words with highest probability. We rescale the mass function to sum up to 1 so that $\reviewerwords$ forms a valid distribution.

To select $\reviewer$, we could now simply add predictive words sampled from this distribution. However, as described earlier, naively doing this will likely have unwanted side effects because of concurring reviewers.
To account for this, we further refine this distribution and simultaneously consider multiple reviewers. Let $\tilde{\reviewer}$ be a targeted reviewer and $\surroundingreviewers$ a set of concurring reviewers. We want to restrict $\reviewerwords_{\tilde{\reviewer}}$ to only include words that are predictive for $\tilde{\reviewer}$ but not for any reviewer in $\surroundingreviewers$. Specifically, we define $\reviewerwords^+_{\tilde{\reviewer}, \surroundingreviewers}$ with 
\[
    \reviewerwordsmass^+_{\tilde{\reviewer}, \surroundingreviewers} \colon \vocabulary \rightarrow \mathbb{R}, \quad \word \mapsto
    \begin{cases*}
        \reviewerwordsmass_{\tilde{\reviewer}}(w) &
        if \parbox[t]{5.5cm}{$\reviewerwordsmass_{\tilde{\reviewer}}(w) \neq 0 \wedge\\   
         \forall r \in \surroundingreviewers: \reviewerwordsmass_{\reviewer}(w) = 0$} \\
        0 & otherwise \\
    \end{cases*}    
\]

Subsequently, to form a valid probability mass function, we rescale $\reviewerwordsmass^+_{\tilde{\reviewer}, \surroundingreviewers}$ to sum up to 1. Note for $\surroundingreviewers = \emptyset$ it follows $\reviewerwords^+_{\tilde{\reviewer}, \surroundingreviewers} = \reviewerwords_{\tilde{\reviewer}}$. 
Sampling from $\reviewerwords^+_{\tilde{\reviewer}, \surroundingreviewers}$ only yields words that are predictive for $\tilde{\reviewer}$ but not $\surroundingreviewers$. 
Often we are also interested in the opposite case, i.e., words that are predictive for all reviewer in $\surroundingreviewers$ but not for $\tilde{\reviewer}$ (e.g., when we want to remove words to promote $\tilde{\reviewer}$ in the ranking). Analogous, we define  $\reviewerwords^-_{\tilde{\reviewer}, \surroundingreviewers}$ and write 
\[
    \reviewerwordsmass^-_{\tilde{\reviewer}, \surroundingreviewers} \colon \vocabulary \rightarrow \mathbb{R}, \quad \word \mapsto
    \begin{cases*}
        \frac{1}{|\surroundingreviewers|}\sum_{\reviewer \in \surroundingreviewers}{\reviewerwordsmass_{\reviewer}(w)} &
        if \parbox[t]{5.5cm}{$\reviewerwordsmass_{\tilde{\reviewer}}(w) = 0 \wedge\\   
         \forall r \in \surroundingreviewers: \reviewerwordsmass_{\reviewer}(w) \neq 0$} \\
        0 & otherwise \\
    \end{cases*}
\]

Again, we rescale $\reviewerwordsmass^-_{\tilde{\reviewer}, \surroundingreviewers}$ to sum up to 1. For $\surroundingreviewers = \emptyset$, the distribution $\reviewerwords^-_{\tilde{\reviewer}, \surroundingreviewers}$ is not well defined, as its mass function always evaluates to 0 and we thus set $\reviewerwords^-_{\tilde{\reviewer}, \surroundingreviewers} := \reviewerwords_{\tilde{\reviewer}}$. Figure \ref{fig:reviewer-words} graphically depict this construction.
For reviewer selection, we consider sets of concurring reviewer $\surroundingreviewers$ that are close to $\reviewer$ in the ranking. Specifically, we randomly sample $\nosuccessors$ subsets from
\begin{align*}
\surroundingreviewers \subseteq Pow(\left \{ \reviewer \mid \forall \reviewer \neq \tilde{\reviewer} \in \reviewersset: 0 \leq \text{rank}_{\tilde{\reviewer}} - \text{rank}_{\reviewer} - \revieweroffset \leq \reviewerwindow \right \})
\end{align*}
for a given reviewer window $\reviewerwindow$ with offset $\revieweroffset$. In other words, we exploit locality and focus on reviewer that are either before or close behind $\reviewer$ in the ranking.
For each subset, we create two candidates by (1) adding $\stepsize$ words from $\reviewerwords^+_{\tilde{\reviewer}, \surroundingreviewers}$ respectively (2) remove $\stepsize$ words from $\reviewerwords^-_{\tilde{\reviewer}, \surroundingreviewers}$. 
Reviewer rejection follows analogous with the distributions interchanged and sets sampled from
\begin{align*}
\surroundingreviewers \subseteq Pow(\left \{ \reviewer \mid \forall \reviewer \neq \tilde{\reviewer} \in \reviewersset: -\reviewerwindow \leq \text{rank}_{\tilde{\reviewer}} - \text{rank}_{\reviewer} + \revieweroffset \leq 0 \right \})
\end{align*}
Finally, for multiple target reviewer in $\requestedreviewers$ and $\rejectedreviewers$, we  consider the union of candidates from individual reviewer.

\subsection{Problem Space}
\label{sec:problem-space}

The result of the feature space attack is a modification vector $\modifications \in~\F$ containing the words that have to be modified in the problem space. These words must be incorporated into an actual template PDF file $\inputpdf' \in \Dom$ such that both the semantics and plausibility constraints are satisfied. 
Fortunately, the assignment system obtains a document as input and not the raw text. This provides an adversary with more capabilities and flexibility. She can carefully manipulate the text of her submission as well as exploit weak spots in the text representation or document format.

\begin{table}[t]
\centering
\footnotesize
\caption{\textbf{Problem-space transformations.} Overview of transformations to realize modifications in the problem space. They are grouped by deniability (text, encoding, format) and the capability to add or delete words. For a detailed description, see Appendix~\ref{app:problem-space-transformations}.}
\vspace{-0.1cm}
\resizebox{\columnwidth}{!}{
\begin{tabularx}{\columnwidth}{XXC{0.6cm}C{0.6cm}}
\toprule 
%Type & Transformation &\rotatebox{90}{\head{Add}} & 
%		\rotatebox{90}{\head{Delete}} \\
& & \multicolumn{2}{c}{Modification} \\
\cmidrule{3-4} 
Type & Transformation & Add. & Del. \\
\midrule
\multirow{4}{*}{Text-level} & Reference addition   & \CIRCLE & \Circle   \\
& Synonym & \CIRCLE & \CIRCLE   \\
& Spelling mistake & \Circle & \CIRCLE \\
& Language model & \CIRCLE & \Circle \\
% \midrule
\cmidrule{2-4}  \addlinespace[5pt]
\multirow{1}{*}{Encoding-level} & Homoglyph      & \Circle & \CIRCLE   \\
% \midrule
\cmidrule{2-4} \addlinespace[5pt]
\multirow{1}{*}{Format-level} & Hidden box & \CIRCLE &  \CIRCLE  \\
\bottomrule
\end{tabularx}}
\label{table:problem-space-overview-transformations}
\vspace{-1.5em}
\end{table}

Consequently, we divide the modifications into \emph{text-level}, \emph{encoding-level}, and \emph{format-level} transformations---sorted according to their deniability. Text-level modifications operate on the actual text, so that only targeted modifications are possible. However, the modifications are deniable if the submission raises suspicion during reviewing. Encoding-level and format-level transformations manipulate the text representation and format, respectively, and enable large modifications, but are not deniable once detected.
Table~\ref{table:problem-space-overview-transformations} lists the transformations implemented in our approach. For a detailed overview, we refer the reader to Appendix~\ref{app:problem-space-transformations}.

\paragraph{Text-level transformations.}
We begin with transformations that are based solely on changes to the visible text and applicable to any text format. As such, they cannot be readily recognized without a semantic analysis of the text.

\newcommand{\minipara}[1]{\emph{#1~}}

\minipara{(a) Reference addition.}
As the first transformation, we consider additions to the submission's bibliography. The transformation adds real references that contain the words to be added. As references, we use publications from security conferences and security-related technical reports. Our evaluation demonstrates that this transformation is very effective, while creating plausible and semantics-preserving changes to a paper. However, it introduces side effects, as not only selected words are added, but also parts of the conference names, authors, and titles. This motivates the hybrid search strategy that we outline in Section~\ref{sec:feature-problem-space}.

\minipara{(b) Synonym.}
We develop a transformation that replaces a word with a \emph{synonym}. To enhance the quality of the proposed synonyms, instead of using a general model for the English language~\mbox{\cite[\eg][]{li-19-textbugger, jin-20-bert, ren-19-generating}}, we use a security-domain specific neural embedding that we compute on a collection of 11,770~security papers. Section~\ref{sec:discussion} presents the dataset. This domain-specific model increases the quality of the synonyms, so that this transformation is also difficult to spot.

\minipara{(c) Spelling mistake.}
As a third type of text-level manipulations, we implement a spelling-mistake transformation, which is common for misleading text classifiers~\cite{gao-18-blackbox, liu-20-joint}. Here, we improve on prior work by trying to find typographical errors from a list of common misspellings~\cite{misc-mispellings} instead of introducing arbitrary mistakes.
For example, the suffix \emph{ance} is often confused with \emph{ence}, so that \say{appearance} becomes the unobtrusive misspelling \say{appearence}. If we do not find such errors, we apply a common procedure from the adversarial learning literature: We either swap two adjacent letters or delete a letter in the word~\cite{li-19-textbugger, gao-18-blackbox, liu-20-joint}. 

\minipara{(d) Language model.} Finally, we apply the large-scale unsupervised language model \mbox{OPT}~\cite{zhang-22-opt} to create text containing the words to be added. To generate security-related text, we finetune the model using the corpus of 11,770 security~papers. While the created sentences are not necessarily plausible, this transformation allows us to technically evaluate the possibility that an adversary creates new text to insert words. Given the increasing capabilities of language models, we expect the chances of creating plausible text to rise in the long run. Moreover, we assume that in practice attackers would manually polish the generated text to reduce their detection probability.

\paragraph{Encoding-level transformations.}
As the second class of transformations, we consider manipulations of the text encoding. These manipulations may include the substitution of characters, the application of unicode operations, or changes to the font face and color. For our implementation, we focus on \emph{homoglyph} transformation, inspired by previous works that replaces characters with visually similar counterparts~\cite{li-19-textbugger, eger-19-text}. By replacing a character with a homoglyph, we can remove selected words from the bag-of-words vector used for the topic model. 
Similarly, there are several other strategies for tampering with text encoding~\citep{boucher-22-bad}. Since these manipulations also change only the visual appearance of the text, we consider homoglyphs as a representative example of the class of encoding-level transformations.

\paragraph{Format-level transformations.}
As the third class of transformations, we focus on changes specific to the underlying document format, such as accessibility features, scripting support, and parsing ambiguity \citep{markwood-17-pdf}. As an example of this class of transformations, we consider \emph{hidden boxes} in the PDF format. 
Our transformation relies on accessibility support with the latex package \texttt{accsupp} to define an invisible alternative text in a hidden box associated with a word. The text extractor processes the alternate text, while PDF readers display only the original word. This discrepancy allows an attacker to add words as alternate text. Likewise, she can put an empty alternative text over a word that should be removed.

\paragraph{Improved transformations.}
In addition, we exploit the preprocessing implemented by assignment systems. 
First, we benefit from stemming, so that the transformations only need to add or delete \emph{stems} instead of words. This increases the possibilities to find suitable text manipulations. For example, an attacker can modify the words \emph{attacker} or \emph{attackable} to remove the feature~\emph{attack}, since both are reduced to the same stem.
Second, we exploit the filtering of stop words. The hidden box transformation requires sacrificing a word for defining an alternative text. As stop words are not part of the feature vector, no side effects occur if they are changed.

\subsection{Feature-Problem-Space Attack}
\label{sec:feature-problem-space}

We are now equipped with (i) a strategy to find modifications~\mbox{$\modifications \in \F$} and (ii) transformations \mbox{$\transformation \in \Dom$} to realize $\modifications$ in a paper submission.
The ultimately missing piece is an optimization strategy that brings these two components together. In general, this optimization is responsible for guiding the transformations towards the targeted assignment. 
In the following, we first present the basic principle of our applied strategy and then introduce two practical extensions.

\paragraph{Hybrid optimization strategy.}
Due to the challenges around the problem space and the feature space, we use a strategy that \emph{switches alternately} between \Dom and \F.
Figure~\ref{fig:overview} on page \pageref{fig:overview} schematically illustrates our alternating approach. For an initial submission~\inputpdf, the adversary extracts the features (step~\stepone) and performs a feature-space attack (step~\steptwo and \stepthree). As $\extractor$ is not invertible, the adversary then has to find suitable transformations in the problem space (step~\stepfour) that realize the requested modifications. This leads to a new feature vector in \F (step~\stepfive). However, this vector is shifted due to side effects and limitations of the transformations. Consequently, the adversary continues her search from this new position and repeats the process iteratively until the target is reached or the maximum number of iterations have passed.

We note that side effects are not always negative as assumed by prior work~\cite{pierazzi-20-intriguing}. In our evaluation, for example, we found that the additional words introduced by the reference transformation can further push a reviewer's rank towards the target, since the additional words may also relate to other selected reviewers, for example, due to co-authors or paper titles. However, the impact of side effects is difficult to predict in advance, so that an optimization strategy should be capable of dealing with positive as well as negative side effects.

\paragraph{Constraint mapping $\Dom \rightarrow \F$.}
Our first extension to this hybrid strategy addresses the complexity of problem-space modifications. In practice, not every requested modification from \F can be realized in~\Dom with the implemented transformations due to PDF and \LaTeX{} restrictions.
For example, in \LaTeX{}, homoglpyhs are not usable in the listing environment, while the hidden box is not applicable in captions or section titles. In general, such restrictions are difficult to predict given the large number of possible \LaTeX{} packages.
Instead of solving such shortcomings in the problem space by tediously adjusting the transformations to each special case, we resort to a more generic approach and transfer problem-space constraints back to the feature space. 
The transformers in \Dom first collect words that cannot be handled, which are then blocked from being sampled during candidate generation in~\F.

\paragraph{Surrogate models.}
We introduce a second extension for the black-box scenario. 
In this scenario, the adversary has no access to the victim model. Still, she can leverage public information about the program committee, collect papers from its members, and assemble a dataset similar to the original training data. This allows her to train a \emph{surrogate model} that enables preparing an adversarial paper without access to the assignment system. This strategy has been successfully used for attacks against neural networks~\cite{papernot-16-transferability}.
However, in our case, this strategy is hindered by a problem: LDA models suffer from high variance \cite{agrawal-18-what, mantyla-18-measuring}. Even if the adversary had access to the original data, she would still get different models with varying predictions. This makes it unlikely that an adversarial paper computed for one model transfers to another. 

As a remedy, we propose to use an ensemble of surrogate models to better approximate the space of possible LDA models. 
We run the attack simultaneously for multiple models until being successful against \emph{all} surrogates. To this end, we extend the feature-space attack to multiple target models: (i) we create candidates for each surrogate model independently and consider the union over all surrogates and (ii) we compute the loss as the sum of individual losses over all surrogates. 
Intuitively, this increases the robustness of an adversarial paper and, consequently, improves the success rate that the attack transfers to the unknown victim model. 

\vspace{-0.25em}
\section{Evaluation}
\label{sec:evaluation}

In the following, we evaluate the efficacy of the proposed approach to prepare adversarial papers. To this end, we simulate the automatic paper-reviewer assignment process of a real conference with the full program committee (PC). 
We consider two different scenarios:
First, we demonstrate how a white-box attacker with full-knowledge about the target system can select and reject reviewers for a submission.
Second, we consider a black-box adversary with only limited knowledge and no access to the trained assignment system. We show that such an adversary can generate effective surrogate models by exploiting public knowledge about the conference.
Finally, we verify that the manipulated papers are plausible and preserve the semantics of the text.

\paragraph{Setup.} 
We use Autobid~\cite{misc-autobid} as an open-source realization of the TPMS concept~\cite{charlin-13-toronto}. We simulate an automatic assignment for the \emph{43rd IEEE Symposium on Security and Privacy} with the full PC and set the paper load $\paperload = 5$ (\ie, assign each submission to five reviewers). In contrast to the real conference, we assume a fully automated assignment without load balancing and conflicts (see Section~\ref{sec:discussion}).
As we do not have access to the original submissions, we use the accepted papers as substitutes. In total, we find 32 papers on the arXiv e-Print archive with \LaTeX{} source, which we use for our evaluation.

The PC consists of 165 persons. For each PC member, we construct an archive $\archive_\reviewer$ of papers representative for the person's expertise by crawling their \emph{Google Scholar} profile. We select 20 paper for each reviewer and compile the corpus as the union of these archives. To simulate a black-box scenario, we additionally generate \emph{surrogate corpuses} that overlap with the original data between 0\%  and 100\%. Appendix \ref{app:corpus} describes this process in detail. In all cases, we train Autobid with the default configuration on a given corpus.

For each attack, we perform a grid search on its parameters to realize a reasonable trade-off between efficacy and efficiency. We start by relaxing any constraints on $\modifications$ ($\maxmannorm = \infty$ and $\maxinfnorm = \infty$) and run the attack with at most $\switches = 8$ transitions between the feature space and problem space (see Appendix \ref{app:hyperparameters} for details).
All experiments are performed on a server with 256 GB RAM and two Intel Xeon Gold 5320 CPUs. % Our code is available at \href{https://github.com/rub-syssec/adversarial-papers}{https://github.com/rub-syssec/adversarial-papers}.

\paragraph{Performance measures.} We use three measures to evaluate the attack's performance. First, we consider an adversarial paper $\inputpdf'$ to be \emph{successful} if the constraints from Equation \ref{eq:attack-goal} are fulfilled.
Second, to quantify modifications to the submission, we use two standard measures: $L_1$ and $L_\infty$ norm. Given the modified word counts $\submission' \coloneqq \submission + \modifications $, these are computed as
	\begin{equation}
	   \modificationsmannorm = \sum_{i}{\left| \modifications_i \right|} \text{ and } \modificationsinfnorm = \max_i \left| \modifications_i \right|.
	\end{equation}
$L_1$ is the absolute number of modified words and provides a general overview on the total amount of modifications. Intuitively, we are interested in minimizing $L_1$ to make an attack less suspicious. Similarly, $L_\infty$ is the maximum change in a single dimension (i.e., a single word) and ensures that a single word is not included too frequently. 
Third, we assess the \emph{semantics} and \emph{plausibility} of the manipulated papers in a user study with security researchers.

\begin{table}[b]
\centering
\footnotesize
\vspace{-1em}
\caption{\textbf{Feature-space search.} We compare our attack with two baselines: hill climbing and morphing. For this comparison, we consider three attack objectives: (1) selecting, (2) rejecting, and (3) substituting of reviewers.}
\label{table:feature-space-search-strategy}
\vspace{-0.1cm}
\resizebox{\columnwidth}{!}{
\begin{tabular}{@{}l c >{\color{gray}}l c >{\color{gray}}l c >{\color{gray}}l @{}}
\toprule
& \multicolumn{2}{@{}c}{Selection} 
& \multicolumn{2}{@{}c}{Rejection} 
& \multicolumn{2}{@{}c}{Substitution} \\
\midrule
\multicolumn{4}{@{}l}{\emph{$L_1$ norm}} \\
\rule{0pt}{2ex}%<--- do not remove
\; Our attack       
    & $~704$ & $\times 1.00$ 
    & $1032$ & $\times 1.00$ 
    & $2059$ & $\times 1.00$\\
\rule{0pt}{2ex}%<--- do not remove
\; Hill climbing 
    & $1652$ & $\times 2.35$
    & $2255$ & $\times 2.18$ 
    & $5526$ & $\times 2.68$ \\
\; Morphing 
    & $3059$ & $\times 4.35$
    & - & $\times \phantom{1}\text{-}\phantom{0}$ 
    & - & $\times \phantom{1}\text{-}\phantom{0}$  \\ [0.5ex]  
\midrule 
\multicolumn{4}{@{}l}{\emph{$L_\infty$ norm}} \\
\; Our attack       
    & $17$ & $\times 1.00$ 
    & $43$ & $\times 1.00$ 
    & $62$ & $\times 1.00$\\
\rule{0pt}{2ex}%<--- do not remove
\; Hill climbing
    & $38$ & $\times 2.22$
    & $44$ & $\times 1.02$ 
    & $98$ & $\times 1.58$ \\
\; Morphing 
    & $45$ & $\times 2.63$
    & - & $\times \phantom{1}\text{-}\phantom{0}$ 
    & - & $\times \phantom{1}\text{-}\phantom{0}$  \\ [0.5ex]  
\bottomrule
\end{tabular}}
\end{table}
\subsection{White-box Scenario}

In our first scenario, we focus on a white-box scenario and consider three attack objectives: (1) selection, (2) rejection, and (3) substitution of a reviewer. For these objectives, we focus on reviewers that are already \say{close} to a submission in the assignment system. For example, a paper on binary analysis would raise suspicion if it would get assigned to a reviewer with a cryptography background.

We use the initial assignment scores of the submission as a proxy to simulate this setting. We determine potential reviewers by computing the ranking for the unmodified submission and consider the 10 reviewers with highest scores. To study objective (1), we sample reviewers from the ranks 6--10 and attempt to get them assigned to the submission. Analogously, for objective (2), we select reviewers from the ranks 1--5 and aim at eliminating their assignment. Finally, for objective (3), we first select a reviewer for removal and then a counterpart for selection. We repeat this procedure 100 times with random combinations of papers and reviewers for each objective. Moreover, to account for the high variance of LDA, we train the topic model 8 times and average results in the following.

\paragraph{Feature-space search.} 

We start our evaluation by examining the feature-space search of our attack in detail. For this experiment, we consider format-level transformations that can realize arbitrary changes. Other transformations are evaluated later when we investigate the problem-space side of our attack.

The results of this experiment are presented in Table \ref{table:feature-space-search-strategy} and further detailed in Appendix \ref{app:boxplots}. We observe that our approach is very effective: 99.7\,\% of the attacks finish successfully with a median run-time of 7 minutes.
The number of performed changes shows high variance, ranging between 9 and 22,621 adapted words. Despite this broad range, however, the average manipulation involves only between 704 and 1,032 words for objectives (1) and (2), respectively. For reference, an unmodified submission contains 7,649 words on average, so that the necessary changes for preparing an adversarial paper amount to 9\% and 13\% of the words.  

Among the three objectives, we see a trend that selecting a reviewer is more efficient than rejecting one. Rejected reviewers have---per construction---a high assignment score, and hence share many topics with nearby reviewers. In contrast, for selected reviewers it is easier to determine topics with less side effects. The third scenario, where we both reject and select a reviewer, is naturally the hardest case. Generally, we observe that topic models based on \ac{LDA} are comparatively robust against adversarial noise, in relation to neural networks which can be deceived into a misclassification by changing only a few words~\cite[e.g.,][]{gao-18-blackbox, li-19-textbugger}.

\paragraph{Baseline experiments.} To put these numbers into perspective, we examine two baselines. 
First, we implement a hill climbing approach that directly manipulates the topic vector of a submission (cf. Equation \ref{eq:score}) by sampling words from the topic-word distributions associated with a reviewer.
For the second baseline, we consider an approach that morphs a target submission with papers that already contains the correct topic-word distribution. 
To find these papers, we compute all assignments of the training corpus and identify submissions to which the target reviewer is assigned. We then repeatedly select words from these submissions and expand our adversarial paper until we reach the target. In rare cases, we could not find papers in which the reviewer is highly rated. We exclude such cases from our experiments.

Considering all three objectives, the hill climbing approach shows a lower success rate: Only 92.2\,\% of the papers are successfully manipulated. The failed submissions either reach the maximum number of 1,000 iterations or get stuck in a local minimum. In successful cases, the attacker needs to introduce more than twice as many changes compared to our attack and the median $L_1$ norm increases from 704--2,059 to 1,652--5,526 words. For the morphing baseline, the attack is successful in only 91.1\,\% of the cases and again needs to introduce significantly more words. We find that the median $L_1$ norm increases by a factor of $4.35$ with a maximum of 29,291 modified words for a single attack.

\paragraph{Generalization of attack.} 
To investigate the generalization of our attack, we repeat this experiment for a second real conference. In particular, we simulate the assignment of the \emph{29th USENIX Security Symposium} with 120 reviewers. We consider 24 original submissions and construct targets as before. Results of this experiment are presented in Appendix \ref{app:generalizaton}.
We observe a similar performance across all three objectives, indicating the general applicability of our attack.

\begin{figure}[t]
    \centering
  	\includegraphics[trim=10 10 10 0, clip, width=\columnwidth]{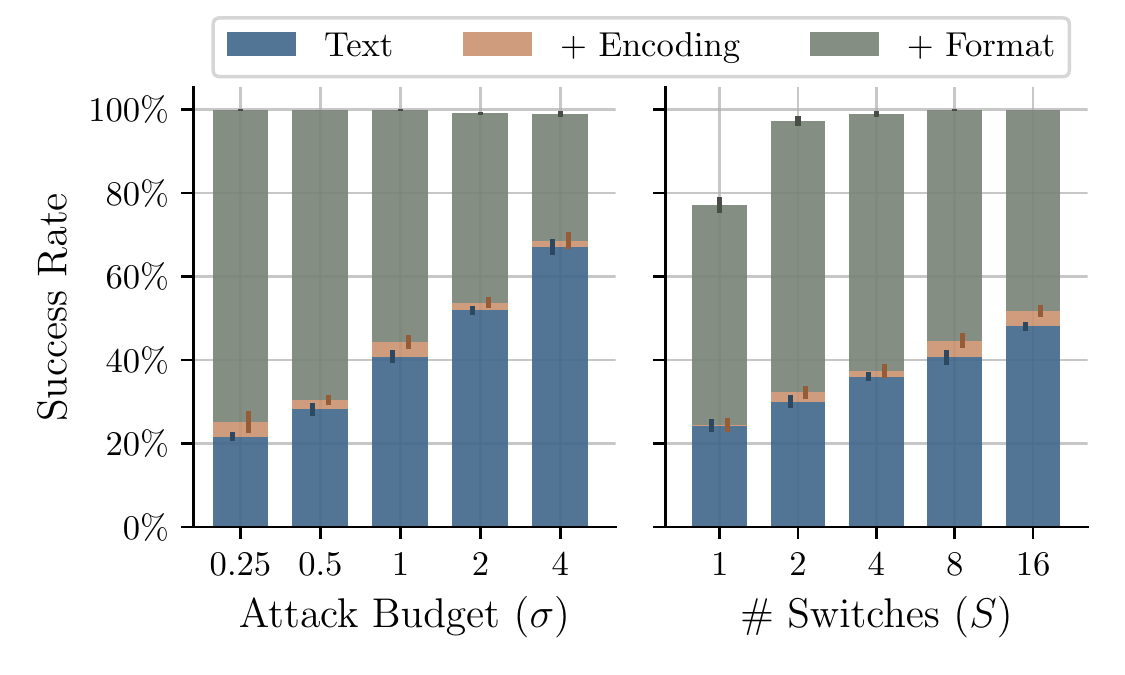}
%   	\vspace{-1.5em}
    \caption{\textbf{Feature-problem-space attack.} We simulate the attack with differently scaled attack budgets $\attackbudgetscale$ (left) and $\switches = 8$ switches. We repeat the experiment (right) with the base budget $\sigma = 1$ and vary $\switches$. For both cases, we randomly select 100 targets from all three objectives that require $\leq 1,000$ changes in $\F$. We report the mean success rate over 8 repetitions.}
    \label{fig:transformer}
\end{figure}
\paragraph{Scaling of target reviewers.}
Next, we scale the attack to larger sets of target reviewers and consider different combinations for selecting, rejecting, and substituting reviewers. We allow an attacker to select up to five target reviewers, which is equivalent to replacing \emph{all} of the initially assigned reviewers. Furthermore, we allow the rejection of up to two reviewers. We focus again on close reviewers and randomly select 100 sets of targets per combination.

The results are summarized in Appendix \ref{app:scaling}. The attack remains effective and we can successfully craft adversarial papers in most of the cases. 
We observe a clear trend that with increasing numbers of target reviewers, we need to perform more changes to the submission. For example, to select all five reviewers, in the median we need to modify 5,968 words. This is expected: we have to move the submission in the topic space from the initially-assigned reviewers to the targeted ones. By adding more reviewers, we include more constraints which results in a significant amount of modifications.

\paragraph{All transformations.}
So far, we have focused on format-level transformations to realize manipulations. These transformations exploit intrinsics of the submission format, which effectively allows us to make arbitrary changes to a PDF file. An attacker likely has access to similar transformations in any practical setting. In fact, robust parsing of PDF files has been shown to be a hard problem \citep[e.g.,][]{carmony-16-extract}. However, we believe it is important for an attacker to minimize any traces and consider different classes of transformations as introduced in Section \ref{sec:problem-space}.

\minipara{(a) Attack budget.}
For this experiment, we introduce an attack budget to describe the maximum amount of allowed modifications for a given transformation. This budget trades off the ability of a transformation to introduce changes with their conspicuousness, since too many (visible) modifications will likely lead to a rejected submission. In particular, we assume a maximum of 25 real and 5 adaptive added \BibTeX{} entries, at most 25 replacements of words with synonyms, no more than 20 spelling mistakes, and up to 10 requested words on average through a text from a \mbox{language model}. In Section~\ref{sec:user-study}, we validate these parameters and assess if the resulting adversarial papers are unobtrusive to human observers.

As a result of the attack budget, we cannot realize arbitrary modifications, since their total amount is restricted. To study this in more detail, we consider the success rate as a function of the attack budget scaled with a factor $\attackbudgetscale$ between 0.25 and 4. During the attack, we split the budget equally across 8 feature-problem-space transitions. We require that targets are feasible with this budget and randomly select 100 targets from the three attack objectives that require $\leq 1,000$ changes in $\F$. Finally, we consider three different configurations: (1) text-level transformations, (2) text-level and encoding-level transformations, and (3) text-level, encoding-level, and format-level transformations combined. We do not restrict the budget for format-level transformations as these transformations are generally not visible.

The results are shown on the left side of Figure \ref{fig:transformer}. For text-level transformations and text-level \& encoding-level transformations, we see an increase in the success rate when the attack budget grows. For the base budget ($\sigma=1$), 40.75\% of the adversarial papers can be prepared with text-level transformations only. That is, no changes in the format and encoding are necessary for manipulating the reviewer assignment.
This can be further improved by increasing the budget, for instance, 67.13\% of the papers become adversarial by scaling it to 4. For smaller budgets, however, we observe that there is often not enough capacity to realize the required modifications. Still, using format-level transformations improves the success rate to 100\% in almost all cases. In rare case, we observe that the attack gets stuck in a local minima. Interestingly, this is more likely with larger budgets. In these cases, the attack makes bigger steps per iteration which introduces more side effects. From the perspective of an attacker, this can be resolved by either increasing the number of switches or reducing the budget.

\minipara{(b) Problem-feature-space transitions.}
To better understand the influence of the alternating search on the success rate of our attack, we conduct an additional experiment. In particular, we simulate our attack for different numbers of transitions $\switches \in \{1, 2, 4, 8, 16\}$ between the problem space and the feature space. We consider the same targets as before and set the attack budget to $\attackbudgetscale = 1$.

The results of this experiment are depicted on the right side of Figure \ref{fig:transformer}. Increasing the number of transitions has a significant effect on the success rate. For all configurations, we see a steady improvement when the number of problem-feature-space transitions increases.
Notably, even the format-level transformations \emph{require} multiple transitions in some cases. The success rate increases from 77.13\%---with no transitions---to 100\% when increasing $\switches$. By alternating between $\F$ and $\Dom$ we share constraints between problem and feature space to find modifications that can be realized in the problem space. This further underlines that it is beneficial and in fact necessary to consider both spaces together.

\begin{figure}[t]
    \centering
  	\includegraphics[trim=10 10 10 10, clip, width=\columnwidth]{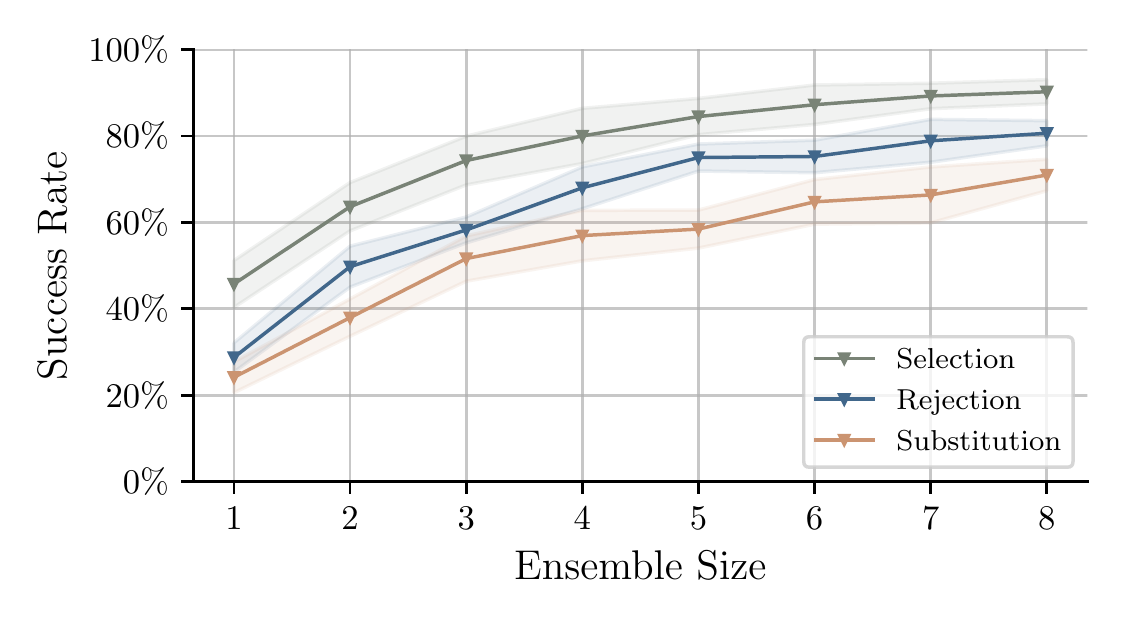}
    \caption{\textbf{Surrogate ensemble sizes.} We simulate the attack with varying numbers of surrogate models. For each ensemble size, we report the mean success rate over 8 target systems with 100 targets each for all three attack objective.}
    \label{fig:surrogates}
\end{figure}
\subsection{Black-box Scenario} 

In practice, an attacker typically does not have unrestricted access to the target system. In the following, we therefore assume a black-box scenario and consider an adversary with only limited knowledge. In particular, this adversary cannot access the assignment system and its training data. Instead, we demonstrate that she could leverage her knowledge about the program committee and construct a surrogate dataset to train her own models for preparing adversarial papers.

The assignment systems AutoBid and TPMS do not specify how the corpus for training a topic model is constructed. They only require that the selected publications are representative of the reviewers. Hence, even if we do not know the exact composition of the training data, we can still collect a surrogate corpus of representative data with public information, such as recent papers of the PC members, and transfer our attack between models. In practice, the success of this transfer depends on two factors: (a) the stability of the surrogate models and (b) the overlap of publications between the original training data and the surrogate corpus.

\paragraph{Stability of surrogate models.}
The training of LDA introduces high variance \cite{agrawal-18-what, mantyla-18-measuring}, so that adversarial papers na\"\i vely computed against one model will likely not transfer to another. To account for this instability, we approximate the model space and consider an \emph{ensemble} of surrogate models.
That is, we run our attack simultaneously against multiple surrogate models trained on the same data. We focus on format-level transformations and repeat the attacks for all three objectives. We vary the number of models in the ensemble from 1 to 8 and consider an overlap of 70\% between the underlying surrogate corpus and the original training data.

Figure~\ref{fig:surrogates} show the results of this experiment. %
Across all objectives, we see an improvement of the success rate when increasing the number of surrogate models. This is intuitive: the adversarial papers are optimized against all models and thus more likely to transfer to other models.
This robustness, however, comes at a cost and the number of modifications increases as well. The median $L_1$ norm increases from 1,990 to 7,556 when considering 8 instead of a single surrogate model (see Appendix \ref{app:surrogate-boxplots}). 

As a result, an adversary in the black-box scenario must find a trade-off: If she needs a successful attack with high probability, she must sacrifice detectability and modify a large number of words. If, on the other end, she only wants to increase her chances for a specific assignment, she can operate without an ensemble and adapt only a few words.

\begin{figure}[t]
    \centering
  	\includegraphics[trim=10 10 10 10, clip, width=\columnwidth]{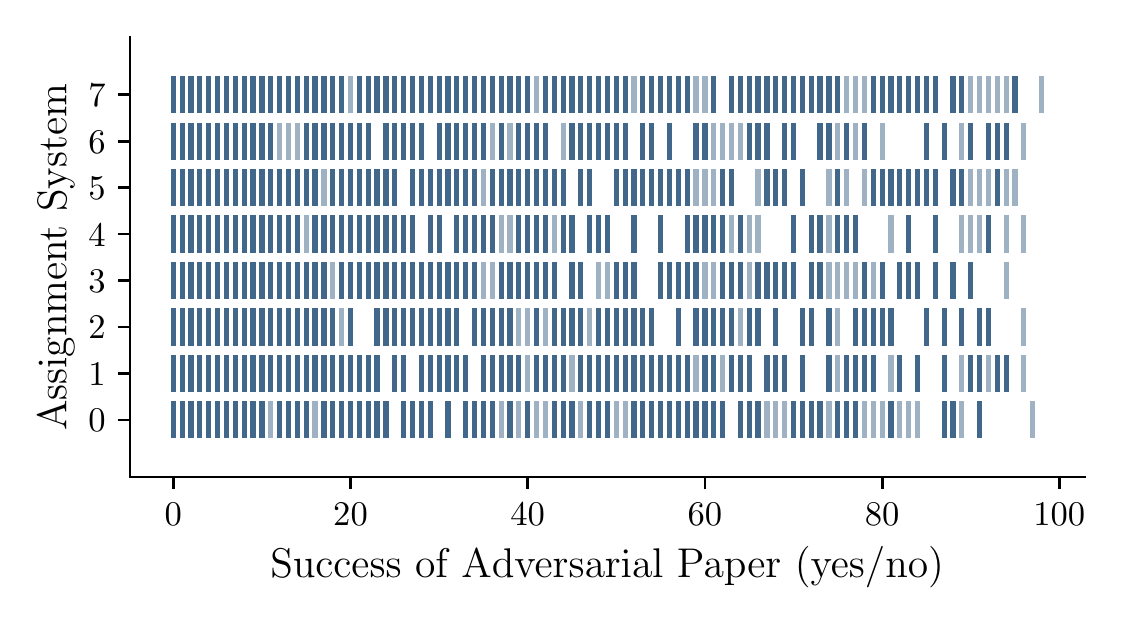}
    \caption{\textbf{Transferability.} We visualize the transferability of 100 adversarial paper among 8 target assignment systems. Attacks were performed with an ensemble size of 8 and we focus on the selection objective. Adversarial papers where the unmodified submission is already successful are displayed in light blue.}
    \label{fig:transferability}
      	\vspace{-1em}
\end{figure}

To further study the transferability of our attack, we sample 100 target reviewer from the median ranking computed over 8 assignment systems and simulate the attack with an ensemble of 8 surrogates. Figure \ref{fig:transferability} visualizes how the resulting adversarial papers transfer among the target systems.
96\% of the papers are successful against four or more target systems and 34\,\% are successful considering all 8 systems.

\paragraph{Overlap of surrogate corpus.}
To understand the role of the surrogate corpus, we finally repeat the previous experiment with varying levels of overlap.
Surprisingly, the attack remains robust against variations in training data. The success rate only fluctuates slightly: 78.0\% (100\% overlap), 80.0\% (70\% overlap), 79.6\% (30\% overlap), and 82.8\% (0\% overlap).
To explain this, we compute the cross-entropy of the reviewer-to-words distributions $\reviewerwords_\reviewer$ for models trained on training data with different overlaps. We observe that the cross-entropy between models trained on the same dataset (i.e., 100\% overlap) is in the same range compared to models trained on different data (cf. Appendix \ref{app:cp-overlap} for details).
As LDA models trained on the same corpus already vary significantly, our attack seeks a robust solution that transfers well if the surrogate models have less overlap with the original training data.

\subsection{Plausibility and Semantics}
\label{sec:user-study}
Finally, we empirically verify if the adversarial modifications are (a) plausible and (b) preserve the semantics of the text. 

\paragraph{Study design.}
As dataset, we use the combined set of original and adversarial papers from our evaluation.
In total, we select seven original papers and their adversarial counterparts, ensuring varying topics and transformations. The attack budget is $\attackbudgetscale = 1.00$. 
Due to a limited number of participants, we focus on visible transformations (i.e.\ encoding-level and text-level) that a reviewer could detect.
Each participant selects (``bids on'') one paper. This selection cannot be changed afterwards and participants are secretly assigned either to the adversarial or to the unmodified version. Each participant will only check one paper to avoid potential bias and fatigue effects.

We design the review process along two phases. Our methodology here is inspired by the work from Bajwa et al. \cite{bajwa-2019-might} and Sullivan et al. \cite{sullivan-10-reviewer}. In the first phase, we request participants to write a mini-review (as a proxy task) for a given paper. In the second phase, we ask if they think the paper has been manipulated. Importantly, the answers of phase~1 cannot be changed. This two-phase separation allows us to observe two factors: First, we can analyze how suspicious adversarial papers are to an unaware reader. Second, once the reader is informed, we can learn about which transformations are noticeable and make our attack detectable.
In each phase, we provide a template with questions on a numerical scale from 1--5, together with free text fields for justifying the rating.
Participants are debriefed finally.
We obtained approval from our institution's Institutional Review Board (IRB) and our study protocol was deemed to comply with all  regulations.

\paragraph{Results.} We recruited 21 security researchers (15$\times$ PhD students, 4$\times$ postdocs, 1$\times$ faculty, 1$\times$ other). All participants are familiar with the academic review process but have different review experience (7$\times$ have not written a review before, $4\times$ between 1-2 reviews, 6$\times$ between 3-10, and 4$\times$ at least 10 reviews). 
The participants reviewed a total of 12 adversarial and 9 original submissions.

\begin{figure}[t]
    \centering
     \includegraphics{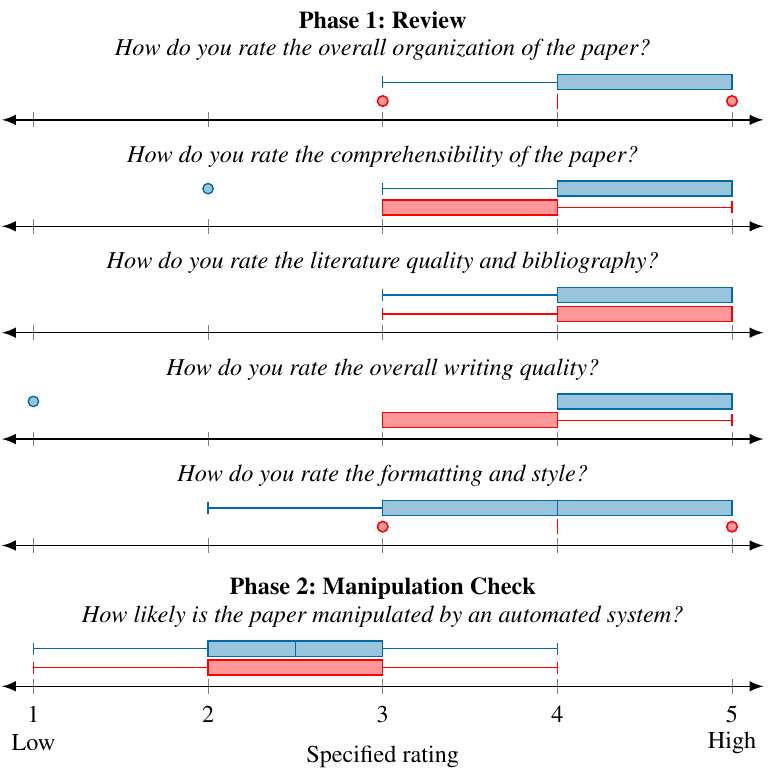}
    \vspace{-0.75em}
    \caption{\textbf{Ratings of benign and adversarial papers.} For each question, the upper boxplot shows the ratings from the benign papers, the lower boxplot from the adversarial papers.}
    \label{fig:user-study-v2}
    % \vspace{-1.5em}
\end{figure}

Figure~\ref{fig:user-study-v2} summarizes the results. Benign and adversarial submissions are rated similar across all review questions. No participant was certain that a paper was manipulated (\ie, gave it a ranking of 5) and only a single of the 12 manipulated submissions was flagged as suspicious with a rating of~4. This was justified with missing references and redundancy in the text---neither of which were introduced by our attack. Interestingly, this reviewer did notice the spelling mistake and language model transformer (when asked about the writing quality), but did not attribute this as a sign for manipulation. This is opposed to two false positive ratings of benign papers, which results in a overall detection precision of 33\% with a recall of only 8\%. This highlights the difficulty to detect any introduced modifications.

Finally, we check that the semantics of the papers are not changed. With the limited attack budget, only small, bounded changes are made to a paper. 
This is further supported by the organization and comprehensibility ratings in Figure \ref{fig:user-study-v2}, which are similar between manipulated and benign submissions.
\section{Discussion}
\label{sec:discussion}

Our work reveals a notable vulnerability in systems for automatic paper-reviewer assignment. 
In the following, we discuss further aspects of our findings, including limitations, defenses, and the implications and benefits of an attack.

\paragraph{Committee size.}
We simulate an automatic assignment for two large security conferences with committees composed of 120 and 165 reviewers, respectively. Considering the current trend, it is likely that these conferences will continue to grow larger. In the following, we want to understand how an increased set of concurring reviewers impacts the attack.
Therefore, we consider committees with 100--500 reviewers sampled from a pool of 528 reviewers (taken from USENIX and S\&P committees between 2018--2022) with a total of 11,770 papers in the underlying corpus.
Appendix \ref{app:committee-size} shows the number of required changes as a function of the committee size. Across the three objectives---selection, rejection, and substitution---we observe only a minor impact on the attack. The attack remains successful with a success rate between 98.00\% -- 98.92\% and the number of required modifications remains largely unaffected. For the smallest committee considered (with 100 reviewers), we observe a slightly larger uncertainty. Intuitively, in this case the assignment is more dependent on the particular choice of the committee which gets averaged out for larger committees.

\paragraph{Load balancing and conflicts.}
Our attack focuses on the manipulation of assignment scores and assumes a direct matching from PC members to submissions. For a complete end-to-end attack, an attacker would also need to take load balancing of submissions and reviewer conflicts into account. For example, the target submission might compete with another paper on the same topic and get a different assignment despite a successful manipulation of the topic model.

Unfortunately, these obstacles are hard to model, as conflicts and the other submissions are typically not known to the adversary. Instead, we can generally improve the resilience of our attack. By increasing the margin $\margin$ of the target reviewer to others, we can make a matching assignment more likely. Interestingly, in this case, conflicts can be even seen as a simplification: if a target reviewer is the top candidate among all reviewers $\reviewersset$, she is also the top candidate for only a subset of reviewers (i.e., all unconflicted reviewers).

To further understand the role of this margin, we simulate the selection of a reviewer for different values of $\margin \in \{0, 0.1, 0.2\}$ and varying numbers of concurring submissions between 200 and 1,000 (sampled from a hold-out corpus). We model the full assignment to maximize similarity subjected to load constraints as introduced in Section \ref{sec:background:assignment}. We assume $\paperload = 5$ reviews per paper and that each reviewer is assigned $\reviewerload = 10$ submissions.
Appendix \ref{app:load-balancing} shows the attack's success rate as a function of the number of concurring submissions. The attack remains effective but we observe a slight downward trend of its success rate. This is expected: with increasing number of submissions, there exist more similar paper that compete for a given reviewer. An attacker can account for this by (1) increasing the margin and, as the attack is undetectable (in general), an attacker could (2) further increase her chances by repeating the attack (e.g., resubmitting a rejected paper).

\paragraph{Paper corpus.}
We select \emph{accepted} papers from IEEE S\&P 2022 as basis for our evaluation. This selection leads to a potential bias, as rejected submissions are not considered. However, we do not expect any impact on our results. Papers follow a common structure, so that our transformations in \LaTeX{} are applicable in general. The feature-space algorithm works on bag-of-words vectors, which is just another representation for any paper.
In Appendix~\ref{app:generalizaton}, we test our attack with papers from the 29th USENIX Security Symposium and find no significant difference in our results.

\paragraph{Countermeasures and defenses.}
Our results show that systems based on topic models such as \ac{LDA} have relatively strong robustness towards adversarial noise. This stands in stark contrast to neural networks, where changing only a few words can already lead to a misclassification~\cite[e.g.,][]{gao-18-blackbox, li-19-textbugger}. However, our work also demonstrates that \ac{LDA}-based systems are still vulnerable to adversarial examples and there is a need for appropriate defenses. 

Unfortunately, text-level manipulations are challenging to fend off, as they can only be spotted on the semantic level. In our user study, participants often struggled to differentiate adversarial modifications from benign issues and an adversary can always \emph{manually} rewrite an adversarial paper to further reduce the detection probability.
Moreover, even completely machine generated text---such as done with our OPT-based transformer---is often indistinguishable from natural text \cite{brown-20-language, mink-22-deepphish}. The underlying models are evolving rapidly and current state-of-the-art models such as 
InstructGPT \cite{ouyang-22-instructgpt} 
and Galactica \cite{taylor-22-galactica} 
are now actively used for academic writing.

For encoding-level and format-level changes, however, defenses are feasible: The root cause of these manipulations is the disparity between human perception and parser-based text extraction. Thus, an effective defense needs to mimic a human reader as close as possible similar to defenses recently proposed for adversarial examples in other domains~\cite[e.g.][]{eisenhofer-21-dompteur}. 
To evaluate this defense, we replace the parser-based text extraction (\texttt{pdftotext}) with an optical character recognition (OCR) (\texttt{tesseract}). We observe that for the modified system the encoding-level and format-level attacks now completely fail, while the performance of the text-level attacks remains unaffected. At the same time, however, we observe a large increase in runtime. Compared to the parser-based extraction, OCR is orders of magnitude slower and needs an average time of 56\,s for a single submission compared to 0.14\,s with conventional text extraction.

Other countermeasures can be more tailored to the individual transformations: Flag usage of unusual font encodings to prevent homoglyph attacks, remove comment boxes and non-typeset pages in a preprocessing step, or automatically verify the bibliography entries using online bibliography databases.

\paragraph{Benefits and implications.}
Manipulating a submission comes with a considerable risk if the attack is detected. This can range from a desk reject over a submission ban at a specific venue to permanent damage of the authors' scientific reputation~\cite{shah-22-challenges}. 
Nevertheless, recent incidents show that academic misconduct happens. Dishonest authors, for example, leveraged synthetic texts to increase the paper output~\cite{cabanac-21-tortured}. Moreover, \emph{collusion rings} exist where authors and reviewers collaborate to accept each other's papers~\cite{littman-21-collusion}. 
Automated assignment techniques can raise the bar for dishonest collaborations considerably~\cite{leyton-22-matching}, yet our work shows that these techniques need to be implemented with care. 
Apart from collusion rings, dishonest authors can also work alone: They can try to promote an unfamiliar reviewer who might overlook paper issues and thus more likely submit a positive review. 

We believe that dishonest authors more likely risk \emph{deniable} manipulations such as a few spelling mistakes or additional references. Our evaluation shows this is sometimes already enough, for example, to promote an unfamiliar reviewer.
As the line between adversarial and benign issues in a paper is often not clear, such an attack can be hard to discover.
All in all, the automatic assignment of papers enables not only manipulations that undermine the entire reviewing process, but also small-scale attacks in which assignments are tweaked by a few deniable changes. 
\section{Related Work}
\label{sec:related}

Our attack touches different areas of security research. In the following, we examine related concepts and methods.

\paragraph{Adversarial learning.} A large body of work has focused on methods for creating adversarial examples that mislead learning-based systems~\cite{biggio-18-wild}. However, most of this work considers attacks in the image domain and assumes a one-to-one mapping between pixels and features. This assumption does not hold in discrete domains, leading to the notion of \emph{problem-space attacks}~\cite{pierazzi-20-intriguing, quiring-19-misleading}. Our work follows this research strand and introduces a new hybrid attack strategy for operating in both the feature space and problem space.
Furthermore, we examine weak spots in \emph{text preprocessing}, which extend the attack surface for adversarial papers. 
These findings complement prior work advocating that the security of preprocessing in machine learning needs be considered in general~\cite{quiring-20-adversarial}. 

Table~\ref{table:related-work-overview} summarizes prior work on misleading text classifiers. While we build on some insights developed in these works, text classification and paper assignment differ in substantial aspects: First, the majority of prior work focuses on untargeted attacks that aim at removing individual features. In our case, however, we have to consider a targeted attack where an adversary needs to specifically change the assignment of reviewers. Second, prior attacks often directly exploit the gradient of neural networks or compute a gradient by using word importance scores. Such gradient-style attacks are not applicable to probabilistic topic models. 

In view of these differences, our work is more related to the attack from \citet{zhou-20-evalda} which studies the manipulation of \ac{LDA}. The authors show that an evasion is \emph{NP-hard} and present an attack to promote and demote individual \ac{LDA} topics. For our manipulation, however, we need to adjust not only individual topics but the complete topic distribution as well as consider side effects with concurring reviewers. 

\begin{table}[t]
\centering
\footnotesize
\caption{Overview of related attacks against text classifiers.}
\vspace{-0.1cm}
\resizebox{\columnwidth}{!}{
\begin{tabularx}{\columnwidth}{p{.125\textwidth}C{0.125cm}C{0.125cm}C{0.125cm}C{0.125cm}C{0.125cm}C{0.15cm}C{0.125cm}C{0.125cm}X}
\toprule 
& \multicolumn{4}{c}{Perturbation} & \multicolumn{2}{c}{Constr.} & \multicolumn{2}{c}{Attack} & \\
\cmidrule(lr){2-5} \cmidrule(lr){6-7} \cmidrule(lr){8-9} 
Paper & \rotatebox{90}{Char} & \rotatebox{90}{Word} & \rotatebox{90}{Sentence}  & \rotatebox{90}{Format} & \rotatebox{90}{Semantics} & \rotatebox{90}{Plausibility} & \rotatebox{90}{Untargeted} & \rotatebox{90}{Targeted} & Classifier \\
\midrule
\emph{This work} & \CIRCLE & \CIRCLE & \CIRCLE & \CIRCLE & \cmark & \cmark & \cmark & \cmark & Assign. \\
\citet{alzantot-18-generating} & \CIRCLE & \CIRCLE & &  & \cmark & \xmark & \cmark & \xmark & NN \\
\citet{ebrahimi-18-hotflip} & \CIRCLE & \CIRCLE & & & \cmark & \xmark & \cmark & \xmark & NN  \\
\citet{eger-19-text} & \CIRCLE & & &  & \cmark & \xmark & \cmark & \xmark  &  NN   \\
\citet{gao-18-blackbox} & \CIRCLE & & &  & \cmark & \xmark & \cmark & \xmark  & NN  \\
\citet{iyyer-18-adversarial} & & & \CIRCLE & & \cmark & & \cmark & \xmark & NN \\
\citet{jin-20-bert} & & \CIRCLE & &  & \cmark & \xmark & \cmark & \xmark  & NN \\
\citet{li-19-textbugger} & \CIRCLE & \CIRCLE & & & \cmark & \xmark & \cmark & \xmark  & NN,LR \\
\citet{liu-20-joint} & \CIRCLE & \CIRCLE & &  & \cmark & \xmark & \cmark & \xmark  & NN \\
\citet{papernot-16-crafting} & & \CIRCLE & & & \xmark & \xmark & \cmark & \xmark  & NN \\
\citet{ren-19-generating} & & \CIRCLE & & & \cmark & \xmark & \cmark & \xmark  & NN \\
\bottomrule
\end{tabularx}}
\label{table:related-work-overview}
% \vspace{-1.5em}
\end{table}
\paragraph{Attacks on assignment systems.}
Finally, another strain of research has explored the robustness of paper-reviewer assignment systems. 
Most of these works are based on \emph{content-masking attacks}~\cite{markwood-17-pdf, tran-19-pdfphantom}, which use format-level transformation to exploit the discrepancy between displayed and extracted text.
More specifically, \citet{markwood-17-pdf} and \citet{tran-19-pdfphantom}, similar to our work, target the paper-reviewer assignment task.
Their attack is evaluated against Latent Semantic Indexing \cite{deerwester-90-indexing}---that is not used in real-world systems like TPMS. 
Although \citet{tran-19-pdfphantom} recognize the shortcomings of format-level transformations, they do not explore text-level transformations or the interplay between the problem space and feature space of topic models. 

Complementary to our work, a further line of research focuses on the collusion of reviewers. These works have analyzed semi-automatic paper matching systems under the assumption that malicious researchers can manipulate the paper assignment by carefully adjusting their paper biddings.
\citet{jecmen-20-mitigating} propose a probabilistic matching to decrease the probability of a malicious reviewer to be assigned to a target submission, while \citet{wu-21-making} tries to limit the disproportional influence of malicious biddings.

\section{Conclusion}
\label{sec:conclusion}

In this paper, we demonstrate that current systems for automatic paper-reviewer assignments are vulnerable and can be misled by \emph{adversarial papers}. On a broader level, we develop a novel framework for constructing adversarial examples in discrete domains through joint optimization in the problem space and feature space. Based on this framework, we can craft objects that satisfy real-world constraints and evade machine-learning models at the same time.

In summary, our work demonstrates a significant attack surface of current matching systems and motivates further security analysis prior to their deployment. As a result, we have informed the developers of TPMS and Autobid about our findings, as part of a responsible disclosure process.

\clearpage
\section*{Acknowledgments}

We thank our shepherd and reviewers for their valuable comments and suggestions. We also thank Ajeeth Kularajan, Andreas Müller, Jonathan Evertz, and Sina Wette for their assistance as well as Charlotte Schwedes and Annabelle Walle for their support with the user study. 
This work was funded by the Deutsche Forschungsgemeinschaft (DFG, German Research Foundation) under Germany's Excellence Strategy -- EXC 2092 CASA -- 390781972), the German Federal Ministry of Education and Research under the grant BIFOLD23B, and
the European Research Council (ERC) under the consolidator grants MALFOY (101043410) and RS$^3$ (101045669).
Moreover, this work was supported by a fellowship within the IFI program of the German Academic Exchange Service (DAAD) funded by the Federal Ministry of Education and Research (BMBF).

\bibliographystyle{abbrvnat}

% squeeze margins in the bib
\let\oldthebibliography\thebibliography
\let\endoldthebibliography\endthebibliography
\renewenvironment{thebibliography}[1]{
  \begin{oldthebibliography}{#1}
    \setlength{\itemsep}{0.0em}
    \setlength{\parskip}{0.0em}
}
{
  \end{oldthebibliography}
}

{\footnotesize\bibliography{strings, references}}

\begin{thebibliography}{59}
\providecommand{\natexlab}[1]{#1}
\providecommand{\url}[1]{\texttt{#1}}
\expandafter\ifx\csname urlstyle\endcsname\relax
  \providecommand{\doi}[1]{doi: #1}\else
  \providecommand{\doi}{doi: \begingroup \urlstyle{rm}\Url}\fi

\bibitem[mis(2021)]{misc-mispellings}
{The Most Common English Misspellings}.
\newblock \href{https://www.lexico.com/grammar/common-misspellings}{Blogpost on
  Lexico}, 2021.

\bibitem[Agrawal et~al.(2018)Agrawal, Fu, and Menzies]{agrawal-18-what}
A.~Agrawal, W.~Fu, and T.~Menzies.
\newblock {What is Wrong with Topic Modeling? And how to Fix it Using
  Search-Based Software Engineering}.
\newblock \emph{Information and Software Technology}, 2018.

\bibitem[Alzantot et~al.(2018)Alzantot, Sharma, Elgohary, Ho, Srivastava, and
  Chang]{alzantot-18-generating}
M.~Alzantot, Y.~Sharma, A.~Elgohary, B.~Ho, M.~B. Srivastava, and K.~Chang.
\newblock {Generating Natural Language Adversarial Examples}.
\newblock In \emph{Conference on Empirical Methods in Natural Language
  Processing}, 2018.

\bibitem[Bajwa et~al.(2019)Bajwa, Langer, K{\"o}nig, and
  Honecker]{bajwa-2019-might}
N.~u.~H. Bajwa, M.~Langer, C.~J. K{\"o}nig, and H.~Honecker.
\newblock {What Might get Published in Management and Applied Psychology?
  Experimentally Manipulating Implicit Expectations of Reviewers Regarding
  Hedges}.
\newblock \emph{Scientometrics}, 2019.

\bibitem[Balzarotti(2020)]{misc-circus}
D.~Balzarotti.
\newblock {System Security Circus}.
\newblock \href{http://s3.eurecom.fr/~balzarot/notes/top4_2020/}{Post on
  personal blog}, 2020.

\bibitem[Bast(2018)]{misc-esa}
H.~Bast.
\newblock {How Objective is Peer Review: The ESA Experiment}.
\newblock
  \href{https://github.com/ad-freiburg/esa2018-experiment/blob/master/BLOGPOST.md}{Blog
  post on CACM}, 2018.

\bibitem[Biggio and Roli(2018)]{biggio-18-wild}
B.~Biggio and F.~Roli.
\newblock {Wild patterns: Ten Years After the Rise of Adversarial Machine
  Learning}.
\newblock \emph{Pattern Recognition}, 2018.

\bibitem[Bird et~al.(2009)Bird, Klein, and Loper]{bird-09-natural}
S.~Bird, E.~Klein, and E.~Loper.
\newblock \emph{{Natural Language Processing with Python}}.
\newblock O'Reilly, 2009.

\bibitem[Blei et~al.(2002)Blei, Ng, and Jordan]{blei-02-lda}
D.~Blei, A.~Ng, and M.~Jordan.
\newblock {Latent Dirichlet Allocation}.
\newblock In \emph{Advances in Neural Information Processing Systems
  (NeurIPS)}, 2002.

\bibitem[Boucher et~al.(2022)Boucher, Shumailov, Anderson, and
  Papernot]{boucher-22-bad}
N.~Boucher, I.~Shumailov, R.~Anderson, and N.~Papernot.
\newblock {Bad Characters: Imperceptible NLP Attacks}.
\newblock In \emph{IEEE Symposium on Security and Privacy (S\&P)}, 2022.

\bibitem[Brown et~al.(2020)Brown, Mann, Ryder, Subbiah, Kaplan, Dhariwal,
  Neelakantan, Shyam, Sastry, Askell, and other]{brown-20-language}
T.~B. Brown, B.~Mann, N.~Ryder, M.~Subbiah, J.~Kaplan, P.~Dhariwal,
  A.~Neelakantan, P.~Shyam, G.~Sastry, A.~Askell, and other.
\newblock {Language Models are Few-Shot Learners}.
\newblock \emph{Advances in Neural Information Processing Systems (NeurIPS)},
  2020.

\bibitem[Cabanac et~al.(2021)Cabanac, Labb{\'{e}}, and
  Magazinov]{cabanac-21-tortured}
G.~Cabanac, C.~Labb{\'{e}}, and A.~Magazinov.
\newblock {Tortured Phrases: {A} Dubious Writing Style Emerging in Science.
  Evidence of Critical Issues Affecting Established Journals}.
\newblock \emph{Computing Research Repository (CoRR)}, 2021.

\bibitem[Carmony et~al.(2016)Carmony, Hu, Yin, Bhaskar, and
  Zhang]{carmony-16-extract}
C.~Carmony, X.~Hu, H.~Yin, A.~V. Bhaskar, and M.~Zhang.
\newblock {Extract Me If You Can: Abusing PDF Parsers in Malware Detectors}.
\newblock In \emph{Symposium on Network and Distributed System Security
  (NDSS)}, 2016.

\bibitem[Charlin and Zemel(2013)]{charlin-13-toronto}
L.~Charlin and R.~Zemel.
\newblock {The Toronto Paper Matching System: An Automated Paper-Reviewer
  Assignment System}.
\newblock In \emph{International Conference on Machine Learning (ICML)}, 2013.

\bibitem[Chaudhuri et~al.()]{misc-cmt}
S.~Chaudhuri et~al.
\newblock {\href{https://cmt3.research.microsoft.com}{Conference Management
  Toolkit (CMT)}}.

\bibitem[Darling(2011)]{darling-11-theoretical}
W.~M. Darling.
\newblock {A Theoretical and Practical Implementation Tutorial on Topic
  Modeling and Gibbs Sampling}.
\newblock In \emph{Annual Meeting of the Assoc. for Computational Linguistics:
  Human Language Technologies (HLT)}, 2011.

\bibitem[Deerwester et~al.(1990)Deerwester, Dumais, Landauer, Furnas, and
  Harshman]{deerwester-90-indexing}
S.~C. Deerwester, S.~T. Dumais, T.~K. Landauer, G.~W. Furnas, and R.~A.
  Harshman.
\newblock {Indexing by Latent Semantic Analysis}.
\newblock \emph{Journal of the American Society for Information Science}, 1990.

\bibitem[Ebrahimi et~al.(2018)Ebrahimi, Rao, Lowd, and
  Dou]{ebrahimi-18-hotflip}
J.~Ebrahimi, A.~Rao, D.~Lowd, and D.~Dou.
\newblock {{H}ot{F}lip: White-Box Adversarial Examples for Text
  Classification}.
\newblock In \emph{Annual Meeting of the Assoc. for Computational Linguistics
  ({ACL})}, 2018.

\bibitem[Eger et~al.(2019)Eger, Sahin, R{\"{u}}ckl{\'{e}}, Lee, Schulz, Mesgar,
  Swarnkar, Simpson, and Gurevych]{eger-19-text}
S.~Eger, G.~G. Sahin, A.~R{\"{u}}ckl{\'{e}}, J.~Lee, C.~Schulz, M.~Mesgar,
  K.~Swarnkar, E.~Simpson, and I.~Gurevych.
\newblock {Text Processing Like Humans Do: Visually Attacking and Shielding
  {NLP} Systems}.
\newblock In \emph{Conference of the North American Chapter of the Assoc. for
  Computational Linguistics: Human Language Technologies, (NAACL-HLT)}, 2019.

\bibitem[Eisenhofer et~al.(2021)Eisenhofer, Sch{\"{o}}nherr, Frank,
  Speckemeier, Kolossa, and Holz]{eisenhofer-21-dompteur}
T.~Eisenhofer, L.~Sch{\"{o}}nherr, J.~Frank, L.~Speckemeier, D.~Kolossa, and
  T.~Holz.
\newblock {Dompteur: Taming Audio Adversarial Examples}.
\newblock In \emph{USENIX Security Symposium}, 2021.

\bibitem[Gao et~al.(2018)Gao, Lanchantin, Soffa, and Qi]{gao-18-blackbox}
J.~Gao, J.~Lanchantin, M.~L. Soffa, and Y.~Qi.
\newblock {Black-Box Generation of Adversarial Text Sequences to Evade Deep
  Learning Classifiers}.
\newblock In \emph{{IEEE} Security and Privacy Workshops (SPW)}, 2018.

\bibitem[Hoffman et~al.(2010)Hoffman, Blei, and Bach]{hoffman-10-online}
M.~D. Hoffman, D.~M. Blei, and F.~R. Bach.
\newblock {Online Learning for Latent Dirichlet Allocation}.
\newblock In \emph{Advances in Neural Information Processing Systems
  (NeurIPS)}, 2010.

\bibitem[Iyyer et~al.(2018)Iyyer, Wieting, Gimpel, and
  Zettlemoyer]{iyyer-18-adversarial}
M.~Iyyer, J.~Wieting, K.~Gimpel, and L.~Zettlemoyer.
\newblock {Adversarial Example Generation with Syntactically Controlled
  Paraphrase Networks}.
\newblock In \emph{Conference of the North American Chapter of the Assoc. for
  Computational Linguistics: Human Language Technologies, (NAACL-HLT)}, 2018.

\bibitem[Jecmen et~al.(2020)Jecmen, Zhang, Liu, Shah, Conitzer, and
  Fang]{jecmen-20-mitigating}
S.~Jecmen, H.~Zhang, R.~Liu, N.~B. Shah, V.~Conitzer, and F.~Fang.
\newblock {Mitigating Manipulation in Peer Review via Randomized Reviewer
  Assignments}.
\newblock In \emph{Advances in Neural Information Processing Systems
  (NeurIPS)}, 2020.

\bibitem[Jin et~al.(2020)Jin, Jin, Zhou, and Szolovits]{jin-20-bert}
D.~Jin, Z.~Jin, J.~T. Zhou, and P.~Szolovits.
\newblock {Is BERT Really Robust? A Strong Baseline for Natural Language Attack
  on Text Classification and Entailment}.
\newblock In \emph{AAAI Conference on Artificial Intelligence (AAAI)}, 2020.

\bibitem[Kohler et~al.()]{misc-hotcrp}
E.~Kohler et~al.
\newblock {\href{https://github.com/kohler/hotcrp}{HotCRP Conference Review
  Software}}.

\bibitem[Lawrence and Cortes(2014)]{misc-nips}
N.~Lawrence and C.~Cortes.
\newblock {The NIPS Experiment}.
\newblock
  \href{https://inverseprobability.com/2014/12/16/the-nips-experiment}{Post on
  personal blog}, 2014.

\bibitem[Leyton{-}Brown et~al.(2022)Leyton{-}Brown, Mausam, Nandwani, Zarkoob,
  Cameron, Newman, and Raghu]{leyton-22-matching}
K.~Leyton{-}Brown, Mausam, Y.~Nandwani, H.~Zarkoob, C.~Cameron, N.~Newman, and
  D.~Raghu.
\newblock {Matching Papers and Reviewers at Large Conferences}.
\newblock \emph{Computing Research Repository (CoRR)}, 2022.

\bibitem[Li et~al.(2019)Li, Ji, Du, Li, and Wang]{li-19-textbugger}
J.~Li, S.~Ji, T.~Du, B.~Li, and T.~Wang.
\newblock {TextBugger: Generating Adversarial Text Against Real-world
  Applications}.
\newblock In \emph{Symposium on Network and Distributed System Security
  (NDSS)}, 2019.

\bibitem[Li and Watanabe(2013)]{li-13-automatic}
X.~Li and T.~Watanabe.
\newblock {Automatic Paper-to-reviewer Assignment, based on the Matching Degree
  of the Reviewers}.
\newblock In \emph{International Conference in Knowledge Based and Intelligent
  Information and Engineering Systems (KES)}, 2013.

\bibitem[Lin et~al.(2020)Lin, Balcan, Hadsell, and Ranzato]{misc-neurips}
H.-T. Lin, M.~F. Balcan, R.~Hadsell, and M.~Ranzato.
\newblock {What we learned from NeurIPS 2020 reviewing process}.
\newblock
  \href{https://neuripsconf.medium.com/what-we-learned-from-neurips-2020-reviewing-process-e24549eea38f}{Blog
  post on Medium}, 2020.

\bibitem[Littman(2021)]{littman-21-collusion}
M.~L. Littman.
\newblock {Collusion Rings Threaten the Integrity of Computer Science
  Research}.
\newblock \emph{Communications of the ACM}, 2021.

\bibitem[Liu et~al.(2020)Liu, Zhang, Wang, Lin, and Chen]{liu-20-joint}
H.~Liu, Y.~Zhang, Y.~Wang, Z.~Lin, and Y.~Chen.
\newblock {Joint Character-Level Word Embedding and Adversarial Stability
  Training to Defend Adversarial Text}.
\newblock In \emph{AAAI Conference on Artificial Intelligence (AAAI)}, 2020.

\bibitem[Liu et~al.(2014)Liu, Suel, and Memon]{liu-14-robust}
X.~Liu, T.~Suel, and N.~Memon.
\newblock {A Robust Model for Paper Reviewer Assignment}.
\newblock In \emph{ACM Conference on Recommender Systems (RecSys)}, 2014.

\bibitem[Long et~al.(2013)Long, Wong, Peng, and Ye]{long-13-good}
C.~Long, R.~C.-W. Wong, Y.~Peng, and L.~Ye.
\newblock {On Good and Fair Paper-Reviewer Assignment}.
\newblock In \emph{IEEE International Conference on Data Mining (ICDM)}, 2013.

\bibitem[Lovins(1968)]{lovins-68-development}
J.~B. Lovins.
\newblock {Development of a Stemming Algorithm}.
\newblock \emph{Mechanical Translation and Computational Linguistics}, 1968.

\bibitem[M{\"{a}}ntyl{\"{a}} et~al.(2018)M{\"{a}}ntyl{\"{a}}, Claes, and
  Farooq]{mantyla-18-measuring}
M.~V. M{\"{a}}ntyl{\"{a}}, M.~Claes, and U.~Farooq.
\newblock {Measuring {LDA} Topic Stability from Clusters of Replicated Runs}.
\newblock In \emph{International Symposium on Empirical Software Engineering
  and Measurement ({ESEM})}, 2018.

\bibitem[Markwood et~al.(2017)Markwood, Shen, Liu, and Lu]{markwood-17-pdf}
I.~Markwood, D.~Shen, Y.~Liu, and Z.~Lu.
\newblock {{PDF} Mirage: Content Masking Attack Against Information-Based
  Online Services}.
\newblock In \emph{USENIX Security Symposium}, 2017.

\bibitem[Mikolov et~al.(2013)Mikolov, Sutskever, Chen, Corrado, and
  Dean]{mikolov-13-distributed}
T.~Mikolov, I.~Sutskever, K.~Chen, G.~S. Corrado, and J.~Dean.
\newblock {Distributed Representations of Words and Phrases and their
  Compositionality}.
\newblock In \emph{Advances in Neural Information Processing Systems
  (NeurIPS)}, 2013.

\bibitem[Mink et~al.(2022)Mink, Luo, Barbosa, Figueira, Wang, and
  Wang]{mink-22-deepphish}
J.~Mink, L.~Luo, N.~M. Barbosa, O.~Figueira, Y.~Wang, and G.~Wang.
\newblock {DeepPhish: Understanding User Trust Towards Artificially Generated
  Profiles in Online Social Networks}.
\newblock In \emph{USENIX Security Symposium}, 2022.

\bibitem[Ouyang et~al.(2022)Ouyang, Wu, Jiang, Almeida, Wainwright, Mishkin,
  Zhang, Agarwal, Slama, Ray, and other]{ouyang-22-instructgpt}
L.~Ouyang, J.~Wu, X.~Jiang, D.~Almeida, C.~L. Wainwright, P.~Mishkin, C.~Zhang,
  S.~Agarwal, K.~Slama, A.~Ray, and other.
\newblock {Training Language Models to Follow Instructions with Human
  Feedback}.
\newblock \emph{Computing Research Repository (CoRR)}, 2022.

\bibitem[Papernot et~al.(2016{\natexlab{a}})Papernot, McDaniel, and
  Goodfellow]{papernot-16-transferability}
N.~Papernot, P.~D. McDaniel, and I.~J. Goodfellow.
\newblock {Transferability in Machine Learning: {F}rom Phenomena to {Black-Box}
  Attacks using Adversarial Samples}.
\newblock \emph{Computing Research Repository (CoRR)}, 2016{\natexlab{a}}.

\bibitem[Papernot et~al.(2016{\natexlab{b}})Papernot, McDaniel, Swami, and
  Harang]{papernot-16-crafting}
N.~Papernot, P.~D. McDaniel, A.~Swami, and R.~E. Harang.
\newblock {Crafting Adversarial Input Sequences for Recurrent Neural Networks}.
\newblock In \emph{{IEEE} Military Communications Conference ({MILCOM})},
  2016{\natexlab{b}}.

\bibitem[Parno()]{misc-autobid}
B.~Parno.
\newblock {Autobid}.
\newblock \href{https://github.com/parno/autobid}{Public GitHub Repository}.

\bibitem[Pierazzi et~al.(2020)Pierazzi, Pendlebury, Cortellazzi, and
  Cavallaro]{pierazzi-20-intriguing}
F.~Pierazzi, F.~Pendlebury, J.~Cortellazzi, and L.~Cavallaro.
\newblock {Intriguing Properties of Adversarial {ML} Attacks in the Problem
  Space}.
\newblock In \emph{IEEE Symposium on Security and Privacy (S\&P)}, 2020.

\bibitem[Quiring et~al.(2019)Quiring, Maier, and Rieck]{quiring-19-misleading}
E.~Quiring, A.~Maier, and K.~Rieck.
\newblock {Misleading Authorship Attribution of Source Code using Adversarial
  Learning}.
\newblock In \emph{USENIX Security Symposium}, 2019.

\bibitem[Quiring et~al.(2020)Quiring, Klein, Arp, Johns, and
  Rieck]{quiring-20-adversarial}
E.~Quiring, D.~Klein, D.~Arp, M.~Johns, and K.~Rieck.
\newblock {Adversarial Preprocessing: Understanding and Preventing
  Image-Scaling Attacks in Machine Learning}.
\newblock In \emph{USENIX Security Symposium}, 2020.

\bibitem[{\v R}eh{\r u}{\v r}ek and Sojka(2010)]{rehurek-10-gensim}
R.~{\v R}eh{\r u}{\v r}ek and P.~Sojka.
\newblock {Software Framework for Topic Modelling with Large Corpora}.
\newblock In \emph{{LREC Workshop on New Challenges for NLP Frameworks}}, 2010.

\bibitem[Ren et~al.(2019)Ren, Deng, He, and Che]{ren-19-generating}
S.~Ren, Y.~Deng, K.~He, and W.~Che.
\newblock {Generating Natural Language Adversarial Examples through Probability
  Weighted Word Saliency}.
\newblock In \emph{Annual Meeting of the Assoc. for Computational Linguistics
  (ACL)}, 2019.

\bibitem[Shah(2022)]{shah-22-challenges}
N.~B. Shah.
\newblock {Challenges, Experiments, and Computational Solutions in Peer
  Review}.
\newblock \emph{Communications of the ACM}, 2022.

\bibitem[Soneji et~al.(2022)Soneji, Kokulu, Rubio-Medrano, Bao, Wang,
  Shoshitaishvili, and Doup{\'e}]{soneji-22-experts}
A.~Soneji, F.~B. Kokulu, C.~Rubio-Medrano, T.~Bao, R.~Wang, Y.~Shoshitaishvili,
  and A.~Doup{\'e}.
\newblock {``Flawed, but like democracy we don’t have a better system'': The
  Experts’ Insights on the Peer Review Process of Evaluating Security
  Papers}.
\newblock In \emph{IEEE Symposium on Security and Privacy (S\&P)}, 2022.

\bibitem[Stelmakh et~al.(2019)Stelmakh, Shah, and
  Singh]{stelmakh-19-peerreview}
I.~Stelmakh, N.~B. Shah, and A.~Singh.
\newblock {PeerReview4All: Fair and Accurate Reviewer Assignment in Peer
  Review}.
\newblock In \emph{Conference on Algorithmic Learning Theory (ALT)}, 2019.

\bibitem[Sullivan et~al.(2010)Sullivan, Baruch, and
  Schepmyer]{sullivan-10-reviewer}
S.~E. Sullivan, Y.~Baruch, and H.~Schepmyer.
\newblock {The Why, What, and How of Reviewer Education: A Human Capital
  Approach}.
\newblock \emph{Journal of Management Education}, 2010.

\bibitem[Taylor(2008)]{taylor-08-optimal}
C.~J. Taylor.
\newblock {On the Optimal Assignment of Conference Papers to Reviewers}.
\newblock Technical report, 2008.

\bibitem[Taylor et~al.(2022)Taylor, Kardas, Cucurull, Scialom, Hartshorn,
  Saravia, Poulton, Kerkez, and Stojnic]{taylor-22-galactica}
R.~Taylor, M.~Kardas, G.~Cucurull, T.~Scialom, A.~Hartshorn, E.~Saravia,
  A.~Poulton, V.~Kerkez, and R.~Stojnic.
\newblock {Galactica: A Large Language Model for Science}.
\newblock \emph{Computing Research Repository (CoRR)}, 2022.

\bibitem[Tran and Jaiswal(2019)]{tran-19-pdfphantom}
D.~Tran and C.~Jaiswal.
\newblock {PDFPhantom: Exploiting {PDF} Attacks Against Academic Conferences'
  Paper Submission Process with Counterattack}.
\newblock In \emph{{IEEE} Annual Ubiquitous Computing, Electronics {\&} Mobile
  Communication Conference ({UEMCON})}, 2019.

\bibitem[Wu et~al.(2021)Wu, Guo, Wu, Kidambi, van~der Maaten, and
  Weinberger]{wu-21-making}
R.~Wu, C.~Guo, F.~Wu, R.~Kidambi, L.~van~der Maaten, and K.~Q. Weinberger.
\newblock {Making Paper Reviewing Robust to Bid Manipulation Attacks}.
\newblock In \emph{International Conference on Machine Learning (ICML)}, 2021.

\bibitem[Zhang et~al.(2022)Zhang, Roller, Goyal, Artetxe, Chen, Chen, Dewan,
  Diab, Li, Lin, and other]{zhang-22-opt}
S.~Zhang, S.~Roller, N.~Goyal, M.~Artetxe, M.~Chen, S.~Chen, C.~Dewan, M.~T.
  Diab, X.~Li, X.~V. Lin, and other.
\newblock {OPT: Open Pre-trained Transformer Language Models}.
\newblock \emph{Computing Research Repository (CoRR)}, 2022.

\bibitem[Zhou et~al.(2021)Zhou, Chen, Zheng, and Wang]{zhou-20-evalda}
Q.~Zhou, H.~Chen, Y.~Zheng, and Z.~Wang.
\newblock {EvaLDA: Efficient Evasion Attacks Towards Latent Dirichlet
  Allocation}.
\newblock In \emph{AAAI Conference on Artificial Intelligence (AAAI)}, 2021.

\end{thebibliography}

\appendix
\section{Training Corpus}
\label{app:corpus}
\vspace{-0.5em}

In the following, we describe how the corpus for the simulated paper-reviewer assignment process is generated. 
The PC of the \emph{43rd IEEE Symposium on Security and Privacy} conference consists of 165 persons. For each PC member, we construct an archive of papers representative for the person's expertise and interests by crawling their \emph{Google Scholar} profile. In rare cases, this profile is not available and we use the profile from \emph{DBLP computer science bibliography} instead. We sort all papers first by year and then by number of citations to obtain an approximation of the recent research interests. From this list, we remove all papers with no citation and for which we cannot obtain a PDF file (e.g., paywalled files we cannot access). Furthermore, we remove papers that are already used as a target submission. From the remaining list, we select the first 40 papers (if available). 

To construct reviewer archives $\archive_\reviewer$, we randomly sample 20 paper for each reviewer and compile the corpus as the union of these archives. The remaining 20 papers are used to simulate the black-box scenario. Here, we consider different levels of overlaps between 0\% (i.e., no overlap between the training data of the surrogates and target system) and 100\% (i.e., complete overlap).

\section{Hyperparameter Selection}
\label{app:hyperparameters}
\vspace{-0.5em}
In the following, we describe how we determine the hyperparameters of our attack in the two scenarios.

\paragraph{White-box scenario.}
We perform a grid search over 100 randomly sampled targets from all three objectives and optimize parameters as a trade-off between attack efficacy and efficiency. To not overfit to a specific model, we train 8 AutoBid systems on different random seeds and randomly select one system per target. Note that an attacker with full-knowledge could also choose parameters that perform best for a specific target. We set the beam width $\beamwidth = 8 \in \{2^0, \dots, 2^3\}$, step size $\stepsize = 128 \in \{2^5, \dots, 2^8\}$, number of successors $\nosuccessors = 512 \in \{2^7,\dots, 2^{9}\}$, and reviewer window to $\reviewerwindow = 6 \in \{2^1, \dots, 2^3\}$ with offset $\revieweroffset = 3 \in \{0, \dots, 3\}$. The target rank for rejected reviewer is set to $rank^{rej}_{\submission} = 10$ and we consider $\reviewerwordsmax = 5000$ words per reviewer. We run the attack for at most $\maxitr = 1000$ iterations with at most $\switches = 8$ transitions between spaces and a target margin of $\margin = -0.02$. 

\paragraph{Black-box scenario.}
We repeat the grid search and train 8 systems on a surrogate corpus at 70\% overlap. We randomly sample 100 targets from all three objectives and assign each a random surrogate system. We set the beam width $\beamwidth = 4 \in \{2^0, \dots, 2^3\}$, step size $\stepsize = 256 \in \{2^5, \dots, 2^8\}$, number of successors $\nosuccessors = 128 \in \{2^7,\dots, 2^{9}\}$, and reviewer window to $\reviewerwindow = 2 \in \{2^1, \dots, 2^3\}$ with offset $\revieweroffset = 1 \in \{0, \dots, 3\}$. Finally, to increase the robustness of our attack, we set the margin as $\margin = -0.16$. All other parameters are the same as before.

\section{Feature-Space Search}
\label{app:boxplots}
\vspace{-0.5em}

We report the $L_1$ norms of individual attacks exemplary for the selection objective. We consider 8 different assignment system and sample 100 random targets per system (i.\,e., 800 attacks in total).\smallskip

\vspace{-1em}
\begin{center}
\includegraphics[trim=10 10 10 10, clip, width=\columnwidth]{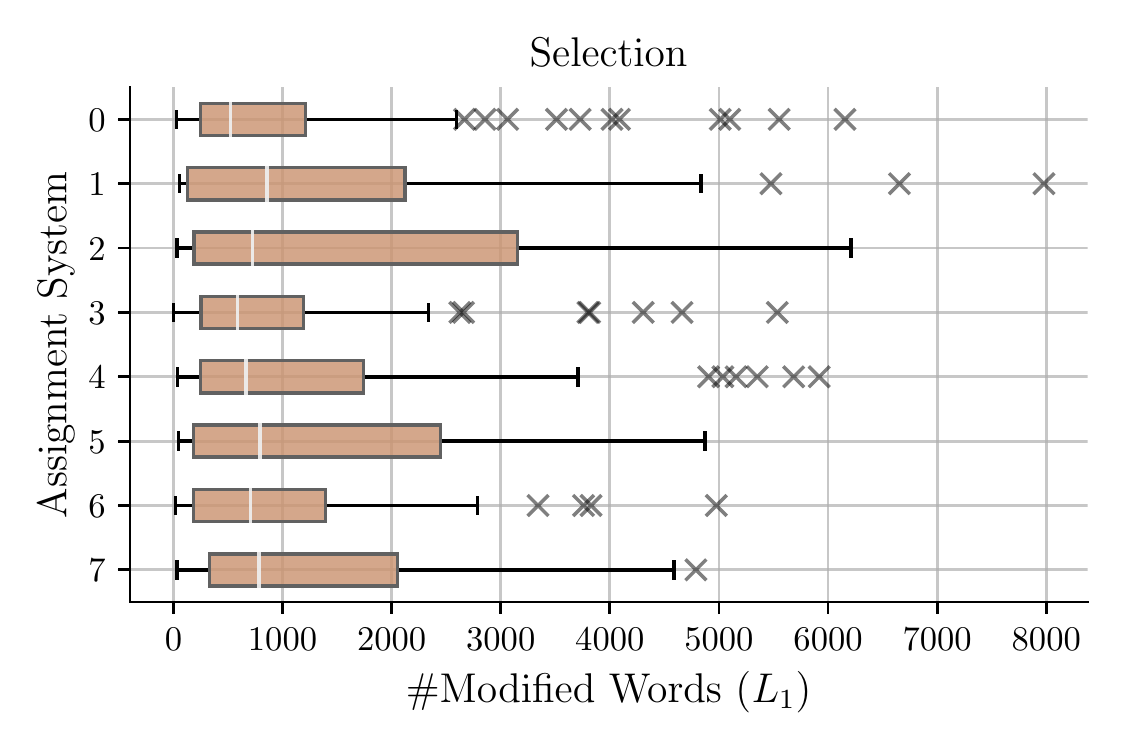}  
\end{center}

\section{Generalization of Attack}
\label{app:generalizaton}
\vspace{-0.5em}

We empirically evaluate our attack on two conferences with differently sized committees: (a) the \emph{29th USENIX Security Symposium} with 120 reviewers and (b) the \emph{43rd IEEE Symposium on Security and Privacy} conference with a larger committee consisting of 165 reviewers. We simulate the attack for all three objectives and report the aggregated success rate, the median running time, and the median $L_1$ and $L_\infty$ norm.

\vspace{-1em}
\begin{center}
\resizebox{1\columnwidth}{!}{
\begin{tabular}{@{}lcccc@{}}
\toprule
& Success Rate & Running Time & \phantom{aaa}$L_1$\phantom{aaa} & \phantom{aaa}$L_\infty$\phantom{aaa} \\
\midrule
USENIX '20 & 99.62\,\% & 7m~38s & $1033$ & $30$ \\
\rule{0pt}{2ex}%<--- do not remove
IEEE S\&P '22 & 99.67\,\% & 7m~12s & $1115$ & $35$ \\
\bottomrule
\end{tabular}}
\end{center}

\section{Scaling of Target Reviewer}
\label{app:scaling}
\vspace{-0.5em}

We run the attack for different combinations of the number of selected and rejected target reviewers. For each combination, we report the median $L_1$ norm as well as the success rate over 100 targets.

\begin{center}
    \includegraphics[trim=10 10 10 22, clip, width=0.95\columnwidth]{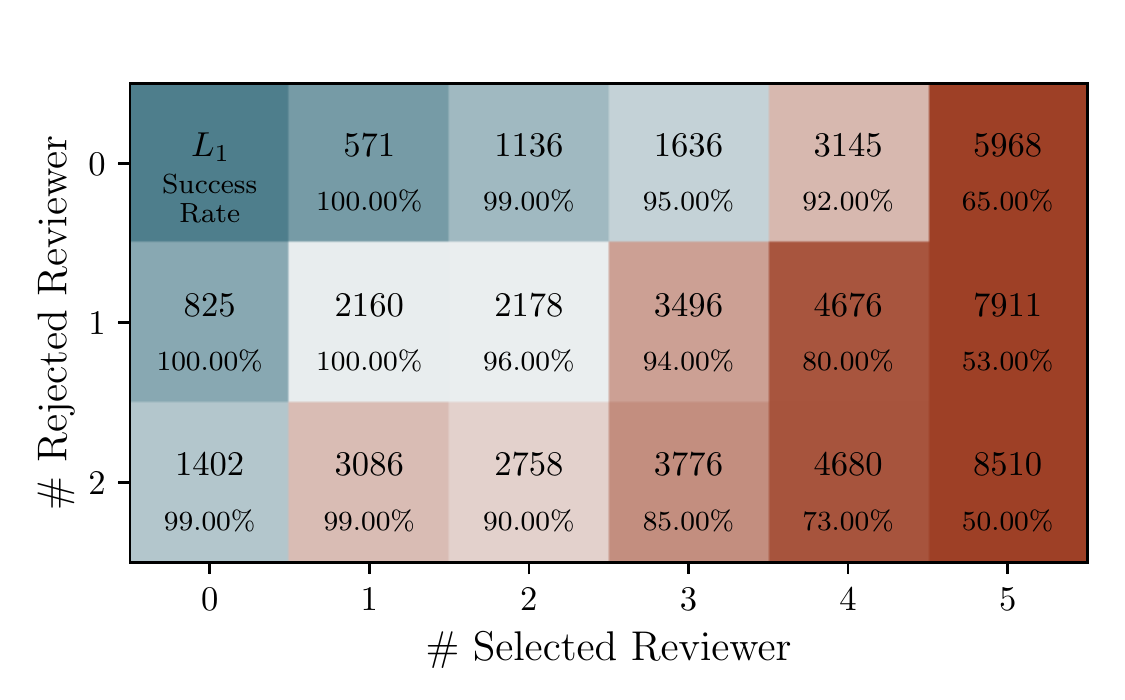}
\end{center}

\newpage
\section{Surrogate Ensembles}
\label{app:surrogate-boxplots}
\vspace{-0.5em}

We report the $L_1$ norms for the black-box scenario with varying sizes of surrogate ensembles. We report the $L_1$ over 100 targets for all three objectives.

\begin{center}
\includegraphics[trim=10 10 10 10, clip, width=\columnwidth]{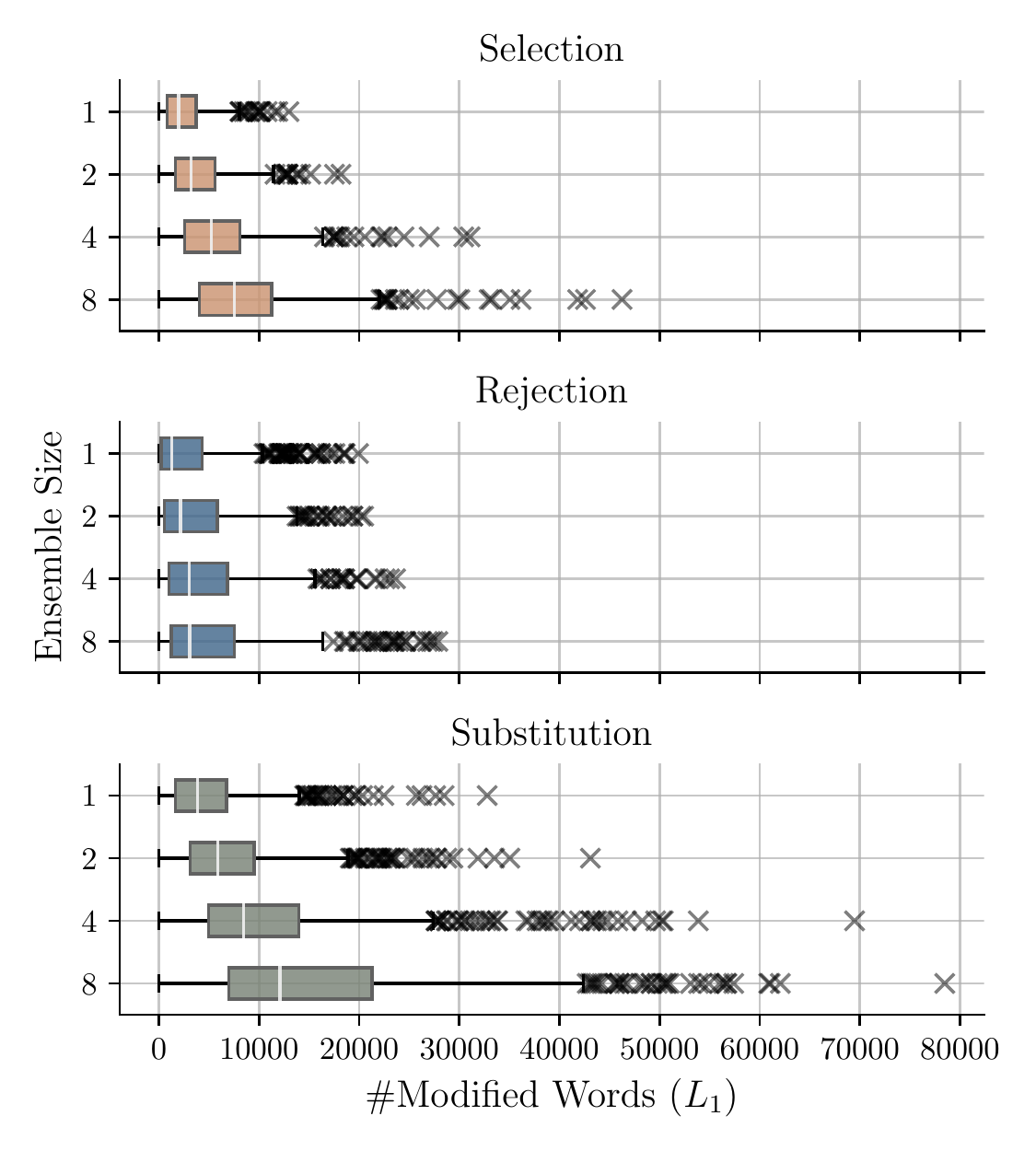}  
\end{center}

\section{Overlap}
\label{app:cp-overlap}
\vspace{-0.5em}

We compare the cross-entropy of reviewer-to-words distributions across models trained on a corpus with different overlaps. We randomly select 10 reviewers and report the mean cross-entropy and standard deviation between 8 models each (i.e., 64 pairs per overlap and reviewer).

\begin{center}
\scriptsize
 \resizebox{0.9\columnwidth}{!}{
        \begin{tabular}{@{}ccccc@{}} 
            \toprule
             \multirow{2}{*}{\#}   & \multicolumn{4}{@{}c}{Overlap} \\
             & 0\% & 30\% & 70\% & 100\%  \\
            \midrule
1 &  $13.19\pm0.46$ &        $13.13\pm0.47$ &        $13.12\pm0.37$ &        $13.20\pm0.44$ \\
% \rule{0pt}{1ex}%<--- do not remove
2 &  $12.56\pm0.29$ &        $12.55\pm0.37$ &        $12.64\pm0.34$ &        $12.50\pm0.29$ \\
% \rule{0pt}{1ex}%<--- do not remove
3 &  $13.58\pm0.63$ &        $13.56\pm0.56$ &        $13.47\pm0.62$ &        $13.52\pm0.63$ \\
% \rule{0pt}{1ex}%<--- do not remove
4 &  $12.43\pm0.50$ &        $12.29\pm0.48$ &        $12.35\pm0.54$ &        $12.32\pm0.50$ \\ 
% \rule{0pt}{1ex}%<--- do not remove
5 &  $13.41\pm0.51$ &        $13.41\pm0.61$ &        $13.50\pm0.56$ &        $13.31\pm0.66$ \\
% \rule{0pt}{1ex}%<--- do not remove
6 &  $12.84\pm0.23$ &        $12.81\pm0.21$ &        $12.93\pm0.25$ &        $12.90\pm0.23$ \\
% \rule{0pt}{1ex}%<--- do not remove
7 &  $14.20\pm0.42$ &        $14.28\pm0.44$ &        $14.39\pm0.48$ &        $14.08\pm0.41$ \\ 
% \rule{0pt}{1ex}%<--- do not remove
8 &  $13.57\pm0.46$ &        $13.59\pm0.46$ &        $13.55\pm0.40$ &        $13.66\pm0.42$ \\ 
% \rule{0pt}{1ex}%<--- do not remove
9 &  $13.44\pm0.72$ &        $13.33\pm0.68$ &        $13.54\pm0.67$ &        $13.44\pm0.76$ \\ 
% \rule{0pt}{1ex}%<--- do not remove
10 &  $15.24\pm0.59$ &        $15.08\pm0.59$ &        $15.31\pm0.66$ &        $14.88\pm0.61$ \\ 
\bottomrule
        \end{tabular}
    }
\end{center}

\newpage
\onecolumn
\enlargethispage{1em}

\begin{multicols}{2}

\section{Committee Size}
\label{app:committee-size}
\vspace{-0.5em}

We simulate the attack with varying sizes of the program committee. For each size, we report the mean number of required modifications over 8 target systems each sampled with a random committee. For each objective, we compute 280 adversarial papers per target system.
\vspace{-0.5em}

\begin{center}
\includegraphics[trim=10 10 10 0, clip, width=\columnwidth]{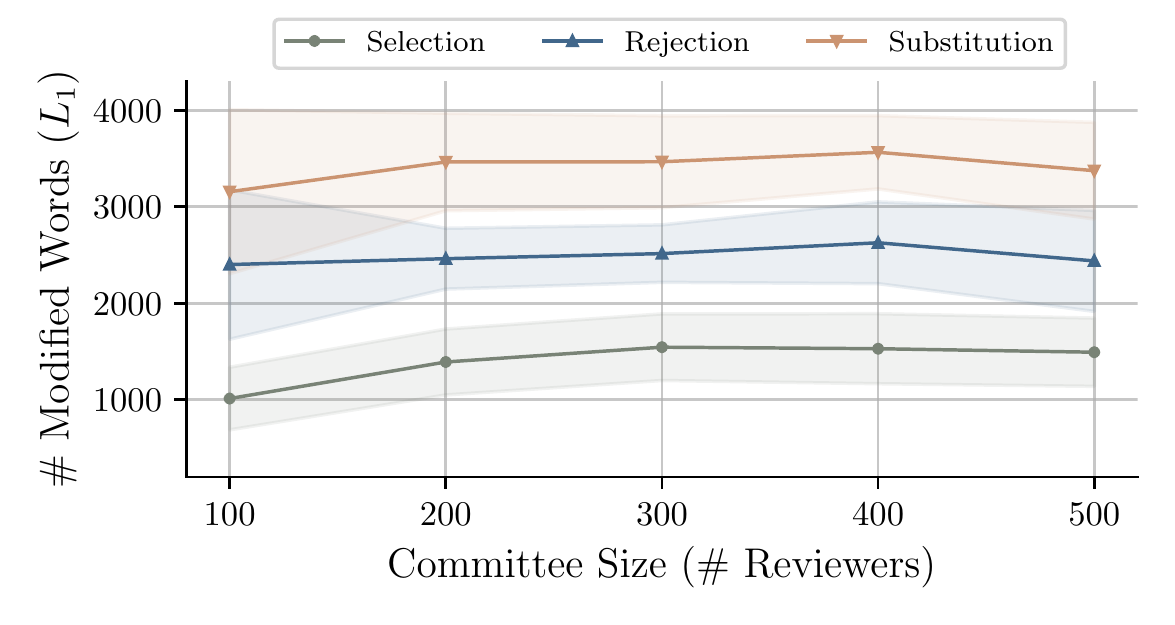}  
\end{center}

\section{Load balancing}
\label{app:load-balancing}
\vspace{-0.5em}

We simulate the attack with varying numbers of concurring submissions between 200 and 1,000. We report the mean success rate over 8 target systems each sampled with a random committee. For each objective, we compute 280 adversarial papers per target system.
\vspace{-0.5em}

\begin{center}
\includegraphics[trim=10 10 10 0, clip, width=\columnwidth]{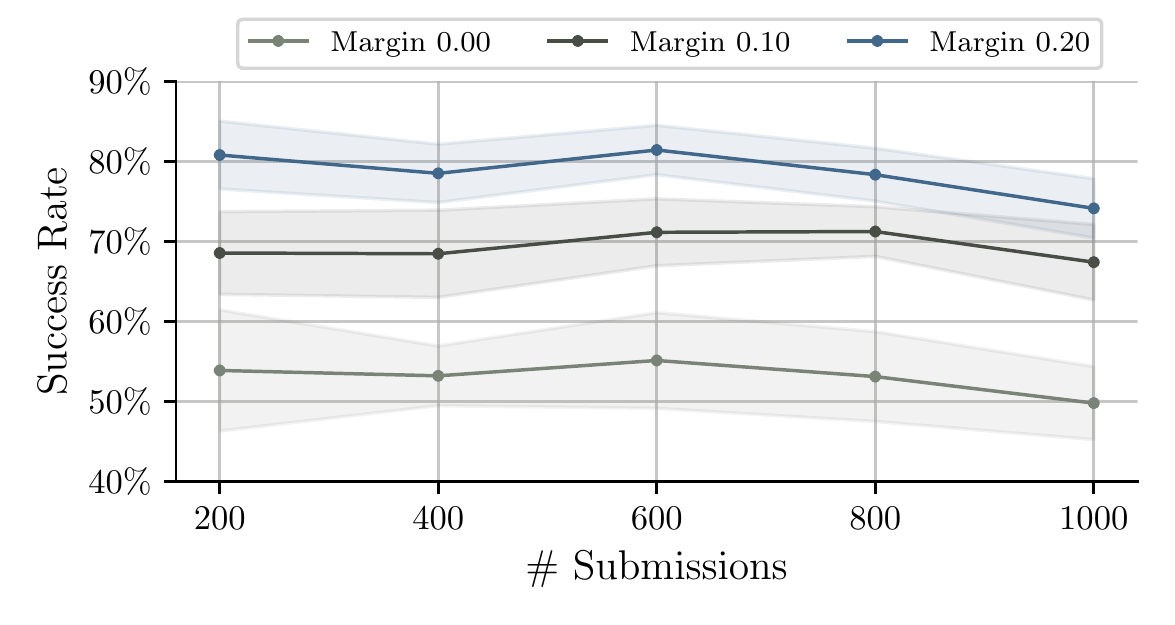}  
\end{center}
\end{multicols}

\vspace{-1.25em}
\section{Problem-Space Transformations}
\label{app:problem-space-transformations}
\vspace{-0.5em}
Detailed description of problem-space transformations. For the transformations' categorization, see Table~\ref{table:problem-space-overview-transformations}.
\vspace{-0.25em}

\newcommand{\head}[1]{#1}

 \begin{center}
 \footnotesize
     \resizebox{1.0\columnwidth}{!}{
	\begin{tabularx}{\textwidth}{p{.135\textwidth}X}
		\toprule
		\head{Transformation} & \head{Description}
		% % % % % % % % % % % % % % % % % %	% % % % % % % % %		
		\\
		\midrule
		\addlinespace[5pt]
	    Reference addition &
		Given a database of bibtex records of real papers, this transformation adds papers to the bibliography.
		It has two options:
		\begin{itemize}
		    \item \emph{Add unmodified papers.} This option treats the insertion as optimization problem. It tries to find $k$ bibtex records such that the number of added words is maximized. This allows us to maximize the impact of the added papers in the bibliography.
		    \item \emph{Add adapted papers.} This option adds $r$ words into a randomly selected bibtex record, which is then added to the bibliography. This transformation allows us to add very specific words which are difficult to add in the normal text in a meaningful way. In the experiments, $r$ is set to 3, \ie, each added bibliography entry has only 3 additional words to avoid suspicion.
		\end{itemize}
		\\ %
		Synonym &
		This transformation replaces words by synonyms using a security domain-specific word embedding~\cite{mikolov-13-distributed}. To this end, the word embeddings are computed on a collection of 11,770 security papers (Section~\ref{sec:discussion} presents the dataset). Two options are implemented:
		\begin{itemize}
		    \item \emph{Add.} Allows adding a word. For each word in the text, it obtains its synonyms. If one of the synonyms is in the list of words that should be added, the synonym is used as replacement for the text word.
		    \item \emph{Delete.} Allows removing a word by replacing it with one of its synonyms.
		\end{itemize}
		The transformation iterates over possible synonyms and only uses a synonym if it has the same part-of-speech (POS) tag as the original word. From the remaining set of synonyms, the transformation randomly chooses a candidate. 
		\\ \tabspac
		Spelling mistake &
		Inserts a spelling mistake into a word that should be deleted. 
		\begin{itemize}
		    \item \emph{Most common misspelling.} This option tries to find a misspelling from a list of 78 rules for most common misspellings, such as \emph{appearence} instead of \emph{appearance} (rule: ends with -ance), or \emph{basicly} instead of \emph{basically} (rule: ends with -ally).
		    \item \emph{Swap or delete.} Swap two adjacent letters or delete a letter in the word. Chooses between both ways randomly.
		\end{itemize}
		The transformation first tries to find a common misspelling, and if not possible, it applies the swap-or-delete strategy.
		\\ \tabspac
		Language model & 
        Uses a language model, here \mbox{OPT}~\cite{zhang-22-opt}, to create sentences with the requested words. To create more security-related sentences, we use the corpus from Section~\ref{sec:discussion} consisting of 11,770 security papers to finetune the OPT-350m model. Equipped with this model, the transformer appends new text at the end of the related work or discussion section. To this end, we extract some text before the insertion position and ask the model to complete the text while choosing suitable words from the set of requested words. 
		\\
		\midrule
		\addlinespace[5pt]
		Homoglyph &
		Replaces a single character in a word by a visually identical or similar homoglyph. 
		For instance, we can replace the Latin letter \emph{A} by its Cyrillic equivalent.
		\\ \tabspac
		\midrule
		\addlinespace[5pt]
		Hidden box &
		Uses the accessibility support with the latex package \emph{accsupp} that allows defining an alternative text over an input text. Only the input text is visible, while the feature extractor processes the alternative text. This allows adding an arbitrary number of words as alternative text. As the input text is not processed, we can also delete words or text in this way. Two options are implemented:
		\begin{itemize}
			\item \emph{Add.} Allows adding an arbitrary number of 
			words in the alternative text. This step requires defining
			the alternative text at least over a visible word that is,
			however, not extracted as feature afterwards anymore. To reduce side effects,
			the transformation first checks if the attack requests a word to be reduced.
			If so, it lays the alternative text over this word. Otherwise, a stop word is 
			chosen that would be ignored in the preprocessing stage 
			anyway. The step thus reduces possible side effects. 
		   \item \emph{Delete.} Adds an empty alternative text over the input word that needs to be removed, so that the word is not extracted anymore.\vspace{-0.8em}
		\end{itemize}\\
		\bottomrule
	\end{tabularx}}
 \end{center}

\end{document}